\documentclass[prd,twocolumn,showpacs,superscriptaddress]{revtex4}
\usepackage{epsfig}
\usepackage{mathrsfs}
\usepackage{amsmath}
\usepackage{amssymb}
\usepackage{color}
\usepackage{graphicx}

\begin{document}

\title{Diffuse Galactic gamma--ray flux at very high energy} 

\author{Paolo Lipari}
\email{paolo.lipari@roma1.infn.it}
\affiliation{INFN, Sezione Roma ``Sapienza'',
 piazzale A.Moro 2, 00185 Roma, Italy}

\author{Silvia Vernetto}
\email{vernetto@to.infn.it}
\affiliation{
INAF, Osservatorio Astrofisico di Torino,
 via P.Giuria 1, 10125 Torino, Italy}

\affiliation{
 INFN, Sezione Torino,
 via P.Giuria 1, 10125 Torino, Italy}

\begin{abstract}
 The observation of the diffuse Galactic gamma ray flux is the most powerful tool to
 study cosmic rays in different regions of the Galaxy, because the 
 energy and angular distributions of the photons encode information about
 the density and spectral shape of relativistic particles in the entire Milky Way.
 An open problem of fundamental importance is whether cosmic rays
 in distant regions of the Milky Way have the same spectral shape
 observed at the Earth or not. If the spectral shape of protons and nuclei is equal in
 all the Galaxy, the dominant, hadronic component of the 
 diffuse gamma ray flux must have an angular distribution that, after correcting
 for absorption effects, is energy independent.
 To study experimentally the validity of this factorization of the energy and angular
 dependence of the diffuse flux
 it is necessary to compare observations in a very broad energy range.
 The extension of the observations to energies 
 $E_\gamma \simeq 0.1$--10~PeV is of great interest,
 because it allows the study of the cosmic ray spectra 
 around the feature known as the ``knee''.
 The absorption probability for photons in this energy range
 is not negligible, and distorts the energy and angular distributions of the diffuse flux,
 therefore a precise calculation of the absorption effects is necessary for the
 interpretation of the data. 
 In this work we present predictions of the
 diffuse gamma ray flux at very high energy, constructed under different hypothesis for the
 space dependence of the cosmic ray energy spectra, and discuss the potential of
 the observations for present and future detectors.
\end{abstract}


\maketitle

\section{Introduction}
\label{sec:introduction} 
Direct observations of cosmic rays (CR) near the Earth
only measure the fluxes present in a small region of the Milky Way
in the vicinity of the solar system, however 
it is clearly of fundamental importance to study the spectra also
in other regions of the Galaxy.
The most powerful method to obtain information on
the spectra of relativistic particles in distant regions of the Galaxy
is the study of the diffuse fluxes of $\gamma$'s and $\nu$'s.
Cosmic ray propagating in interstellar space can interact with gas 
or radiation fields, and these inelastic interactions generate
gamma rays and neutrinos. These secondary particles travel along straight lines
and are observable at the Earth, forming fluxes that encode the
space and energy distributions of CR in the entire volume of
the Galaxy. 

An open question of great importance in high energy astrophysics is whether
the CR spectra have the same shape 
in all points of the Galaxy, or there is a non trivial space dependence.
To study this problem experimentally it is
clearly very desirable to measure the diffuse gamma ray and neutrino fluxes
in a very broad energy range.

The total flux of $\gamma$ rays can be naturally decomposed into the sum
of several components:
(i) an ensemble of point--like or quasi--point like sources;
(ii) an isotropic flux of extragalactic origin;
(iii) a diffuse Galactic flux generated by the interactions of cosmic rays
in interstellar space.

In the energy range 0.1--1000~GeV the diffuse flux
is the largest of the three components, and has been
detected and studied by several gamma ray telescopes on satellites.
Measurements of the diffuse flux have been obtained by 
OSO--3 \cite{oso3}, SAS--2 \cite{sas2},
COS--B \cite{cosb} and EGRET \cite{egret}. More recently
the Fermi telescope has obtained accurate measurements of the flux over the entire sky
\cite{Ackermann:2012pya,Acero:2016qlg}.

The diffuse flux is concentrated in a narrow region around the Galactic
equator, with one half of the total in the
latitude range $|b| \lesssim 5^\circ$. The flux is 
also larger toward the Galactic Center,
and the contribution from the longitude region $|\ell| < 90^\circ$
is approximately twice as large as the flux from the opposite hemisphere.

Ground based gamma ray detectors
\cite{Atkins:2005wu,Abdo:2008if,Bartoli:2015era,Abramowski:2014vox} 
have also obtained measurements of the diffuse $\gamma$ ray flux in the TeV energy range
for some regions near the Galactic equator.
At higher energy ($E \gtrsim 20$~TeV) there are at present only upper limits
for the diffuse flux \cite{Chantell:1997gs,Borione:1997fy,Apel:2017ocm},
however new detectors (such as IceTop and LHAASO) have the sensitivity
to extend the observations to the PeV energy range. 
These measurements of the diffuse Galactic gamma ray flux at very high energy
can be very important in the study of a possible space dependence of the CR spectra.

Recently the IceCube detector
\cite{Aartsen:2013jdh,Aartsen:2014gkd,Aartsen:2015rwa,Aartsen:2017mau}.
has obtained evidence for the existence of an astrophysical signal
of very high energy neutrinos ($E_\nu \gtrsim 30$~TeV).
The simplest interpretation of the data is that most of the signal
is of extragalactic origin, however it is also possible that there
is a subdominant contribution due to Galactic emission.
Some authors have also speculated that most, or even all
of the IceCube signal is of Galactic origin.
This however requires a non--standard emission mechanism,
because the angular distribution of the neutrino events that form the signal,
in contrast to gamma ray observations at lower energy,
is approximately isotropic.

If the neutrinos are generated by the standard mechanism of production,
the $\nu$ emission is accompanied by
a $\gamma$ emission of similar intensity and spectral shape.
In the energy range $E \gtrsim 20$~TeV, photons traveling over extragalactic
distances are completely absorbed by interactions with low energy radiation fields.
On the other hand the flux of Galactic gamma rays
is suppressed and distorted by absorption effects but remains observable.
The conclusion is that the Galactic component of the IceCube signal should
have an observable gamma ray counterpart.
The comparison of the gamma ray and neutrino fluxes
in the same energy range is therefore very important to clarify
the origin of the IceCube signal.
In this study it is important to
take into account the effects of absorption of the flux of high energy photons,
that depend on the space distribution of the emission
and have non trivial energy and angular dependences.

The goal of this paper is to discuss the potential of
existing and future observations of the Galactic diffuse flux
of gamma rays at very high energy.
In our discussion we will construct and compare
two predictions of the diffuse
flux at very high energy based on two alternative frameworks to
extrapolate the measurements performed at lower energy.
In both cases the dominant source of Galactic diffuse emission
is the so called hadronic mechanism, where the gamma rays are generated
in the inelastic collisions of protons and other nuclei.
In one model we will assume that CR protons and nuclei
have the same spectral shape, equal to what is observed at the Earth,
in all of the Galaxy.
This implies that the energy distribution of the gamma ray emission
has a shape that is independent from position and therefore,
if photon absorption during propagation
is negligible, that the angular distribution of the flux at the Earth has an energy independent shape.
In an alternative model,
following some previous studies \cite{Gaggero:2014xla,Acero:2016qlg,Yang:2016jda},
we will assume that the CR in the inner part of the Galaxy
have a harder spectrum than what is observed at the Earth. 
Accordingly the space and energy dependences of the $\gamma$ emission are
not factorized and the angular shape of the diffuse flux is energy dependent,
with the fraction of the diffuse flux that arrives from the region around
the Galactic Center increasing with energy.

Some ``non--standard'' models for the diffuse Galactic gamma ray flux at very high energy,
recently proposed on the basis of the IceCube results,
and that predict a very different angular distribution,
will be also very briefly discussed.

The work is organized as follows. 
In the next section we review some general properties of the
diffuse Galactic flux of gamma rays.
In section~\ref{sec:qlocal} we compute the ``local''
rate of emission of gamma rays in the vicinity of the solar system.
This calculation requires only a knowledge of
the CR spectra that are directly observable at the Earth.
In section~\ref{sec:hydrogen},
we construct a simple (cylindrically and up--down symmetric) model for the
interstellar gas density that is adequate to describe the main large scale
features of the distribution.
Section~\ref{sec:fermi} contains a brief discussion
of the existing observations, 
in particular those performed by the Fermi telescope.
The following two sections present two models for the extrapolations to high energy of the Fermi measurements.
Section~\ref{sec:icecube} discusses some ``non--standard''
model for the Galactic diffuse flux
motivated by the IceCube results.
The following section very briefly discusses the
problem of the separation of the diffuse
flux from the flux generated by the ensemble of all
discrete Galactic sources.
In section~\ref{sec:observations} we 
review the existing measurements of the
diffuse Galactic flux at high energy and compare the results to our
calculations. The section also discusses
the perspectives and necessary conditions to
extend the measurements of the diffuse flux up to the PeV energy range.
A final section gives a summary and some conclusions.

\section{The diffuse Galactic $\gamma$ ray flux}
\label{sec:diffuse} 

The diffuse flux of gamma rays with energy $E$ from the direction $\Omega$
can be calculated integrating the emission along the line of sight
and including a correction for absorption effects:
\begin{equation}
 \phi_\gamma (E, \Omega) = \frac{1}{4 \, \pi} ~
 \int_0^\infty ~dt ~q_\gamma [E, \vec{x}_\odot + t \, \hat{\Omega}] ~
e^{- \tau (E, \Omega, t )}~.
\label{eq:phi-diffuse}
\end{equation}
In this expression $q_\gamma (E, \vec{x})$ is the emission rate, that is 
number of gamma rays of energy $E$ emitted per unit volume, unit time and unit energy
from the point $\vec{x}$. The integration is over all points along the line of sight,
with $\vec{x}_\odot$ the position of the solar system,
and $\hat{\Omega}$ the versor in the direction $\Omega$. The factor 1/($4 \, \pi$)
follows from the assumption that the emission is isotropic.
The exponential factor in Eq.~(\ref{eq:phi-diffuse})
gives the survival probability of gamma rays during propagation, and 
$\tau(E, \Omega, t)$ is the optical depth.

The dominant mechanism for gamma ray emission at high energy
is the so called ``hadronic mechanism'', that
is the creation and decay of unstable mesons
(mostly $\pi^\circ$ with smaller contribution of other particles
such as $\eta$ and $\eta^\prime$)
in the inelastic interactions of protons and other CR nuclei with interstellar gas.
The largest contribution to the hadronic emission is due to $pp$ interactions
between relativistic protons and the hydrogen component 
of the interstellar gas. This contribution can be calculated as:
\begin{eqnarray}
 &~ & q_\gamma^{(pp)} (E_\gamma, \vec{x}) = 
 4 \, \pi ~n (\vec{x}) ~~ \times ~~~~~~~~~~~~~~~~~~~~~~~~~~~~ \nonumber \\[0.22cm]
 & ~&
~ \times \int_{E_\gamma}^\infty dE_p
 ~ \phi_p(E_p, \vec{x})
 ~\sigma_{pp} (E_p) 
 \frac{dN_{pp \to \gamma}}{dE_\gamma} (E_\gamma, E_p) ~~~~~~~~ 
\label{eq:q-pp}
\end{eqnarray}
where $n(\vec{x})$ is the number density of hydrogen gas
at the point $\vec{x}$, $\phi_p(E_p, \vec{x})$ 
is the flux of CR protons with energy $E_p$ at the same point, 
$\sigma_{pp} (E_p)$ is the inelastic $pp$ cross section
and $dN_{pp \to \gamma}/dE_\gamma (E_\gamma, E_p)$ is the inclusive spectrum of 
gamma rays generated in a $pp$ interaction after the decay of 
all unstable particles created in the collision.
The integration is over all proton energies $E_p$
that can generate photons with energy $E_\gamma$.
Interactions where the projectile and/or the target is a nucleus
(such as $p$--helium, helium--$p$, helium--helium, and so on) also contribute
to the hadronic emission and can be calculated with expressions 
that have the same structure as Eq.~(\ref{eq:q-pp}) with obvious substitutions.

Smaller contributions to the gamma ray emission are generated by leptonic processes
where the gamma rays are radiated by CR electrons and positrons, via
bremsstrahlung and Compton scattering.
For bremsstrahlung (interactions such as $e + Z \to e + Z + \gamma$)
the target, as in the hadronic emission case,
is interstellar gas. 
For Compton scattering ($e + \gamma_{\rm soft} \to e + \gamma$)
the target is the ensemble of the soft photons that
form the radiation fields in space. In this case it is
necessary to model not only the density, but also the 
energy spectrum and angular distribution of the target particles.

\subsection{Gamma ray absorption}
\label{sec:absorption}
The most important process that can absorb photons during propagation
in interstellar space is pair production interactions
($\gamma \gamma \to e^+ e^-$) where high energy gamma rays interact with
the soft photons that form the Galaxy radiation fields.
The interaction probability per unit length $K(E, \hat{p}, \vec{x}$)
for a photon of energy $E$ and direction $\hat{p}$ at the space point $\vec{x}$
can be calculated, integrating over the
energy and angular distributions of the target photons:
\begin{equation}
 K (E, \hat{p}, \vec{x}) =
 \int d^3 \kappa
 ~ (1- \cos \theta_{\gamma \gamma}) ~n_\gamma (\vec{\kappa}, \vec{x})
 ~\sigma_{\gamma \gamma}(s)
\label{eq:k-absorption}
\end{equation}
In this expression 
$\vec{\kappa}$ is the 3--momentum and $\varepsilon = |\vec{\kappa}|$
is the energy of the target photon, 
$\cos\theta_{\gamma\gamma} = \hat{p}\cdot \hat{\kappa}$
is the cosine of the angle between the interacting particles, and
$\sigma_{\gamma\gamma} (s)$ is the pair production cross section,
that can be expressed as a function of the square of the center of mass energy
$s = 2 E \varepsilon (1- \cos \theta_{\gamma \gamma})$.

The optical depth $\tau(E, \Omega,t)$
for photons observed at the Earth with energy $E$, direction $\Omega$
that have traveled a distance $t$,
can be calculated integrating the interaction probability along the photon trajectory:
\begin{equation}
\tau (E, \Omega, t) = 
\int_0^{t} dt^\prime ~K(E, -\hat{\Omega}, \vec{x}_\odot + t^\prime \; \hat {\Omega})~.
\end{equation}

\begin{figure}[hbt]
\begin{center}
\includegraphics[width=8.0cm]{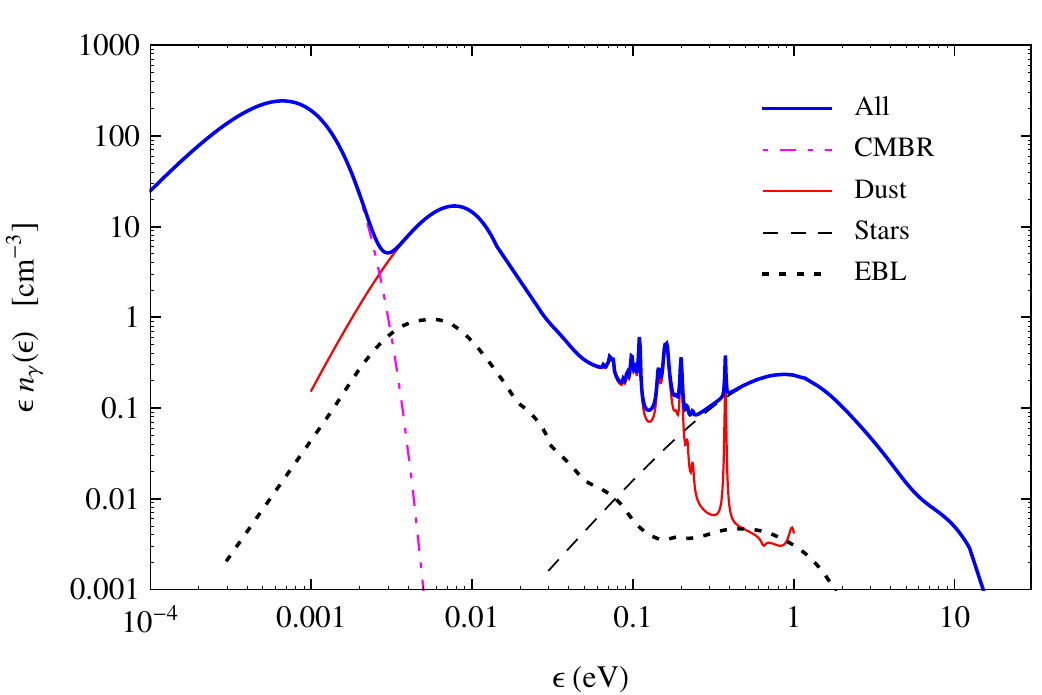}
\vspace{0.3 cm}
\includegraphics[width=8.0cm]{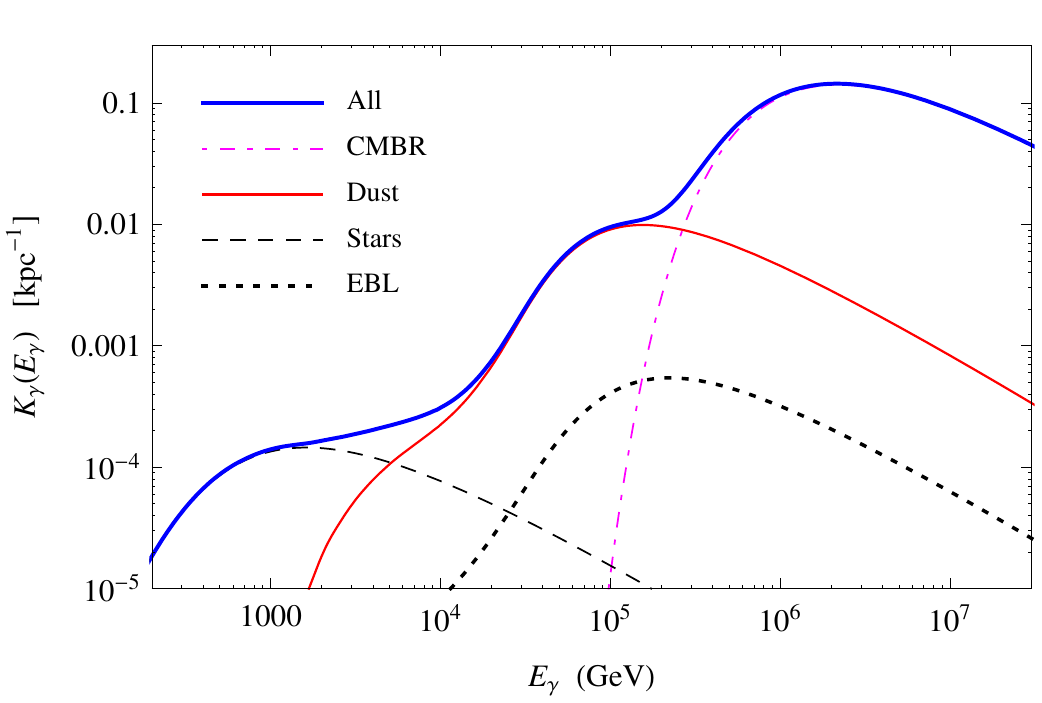}
\end{center}
\caption {\footnotesize
Top panel: energy distribution of soft photons in the solar neighborhood.
Bottom panel: absorption coefficient $K(E, \vec{x}_\odot)$ 
for gamma--rays in the solar neighborhood, as a function
of energy, averaged over the direction of the photon.
In both panels the curves are given for the total and for the
contributions of each component of the target radiation field.
\label{fig:k-abs}}
\end{figure}

The calculation of the absorption probability $K(E, \hat{p}, \vec{x})$ and
of the optical depth $\tau(E, \Omega, t)$ requires a sufficiently accurate knowledge
of the energy and angular distribution of the target photons.
An extended discussion of this problem is contained in \cite{Vernetto:2016alq}
(see also \cite{moska2006} and references therein).
Some of the main properties of high energy gamma ray absorption are illustrated
in Fig.~\ref{fig:k-abs}.
The top panel of the figure shows the (angle integrated)
energy distributions of the target photons in the vicinity
of the solar system according to \cite{Vernetto:2016alq}. The distribution is the superposition of three main components:
the cosmic microwave background radiation (CMBR), infrared emission, and stellar light. A minor contribution is given by the Extragalactic Background Light (EBL).

The CMBR fills homogeneously all space with an isotropic
blackbody spectrum of temperature $T_{\rm CMBR} = 2.7255$~Kelvin;
this corresponds to a total number density $N_\gamma \simeq 410.7$~cm$^{-3}$
of photons with average energy $\langle \varepsilon \rangle \simeq 6.3 \times 10^{-4}$~eV.

Infrared photons are radiated by interstellar dust heated by 
stellar light. This emission can be reasonably well described as a diluted,
and distorted black body spectrum
[$n_\gamma (\varepsilon) \propto n_\gamma^{\rm bb} (\varepsilon, T) \; \varepsilon^{-\beta}$]
with a temperature $T$ of approximately 20~Kelvin and a
distortion parameter $\beta$ of order 1.5--1.7.
At high energy ($\varepsilon \gtrsim 0.03$~eV) the spectral shape deviates from this form 
because of the contribution of an ensemble of emission lines radiated by
the smallest dust grains that are not in thermal equilibrium.
The infrared radiation has an average energy of order 0.008~eV,
and a number density $\simeq 25$~cm$^{-3}$.

Stellar light can be described as the superposition of
diluted black body spectra with temperatures
between 3000 and 8000 Kelvin, plus a small contribution in the ultraviolet range
from young hot stars. In the vicinity of the solar system
the stellar light radiation field has a total
number density of order 0.5~cm$^{-3}$
of photons with average energy $\langle \varepsilon \rangle \simeq 1$~eV.
The infrared and stellar light components of the radiation field
have non trivial space and angular distributions that
reflect the disk structure of the Galactic sources.

The bottom panel of Fig.~\ref{fig:k-abs}
shows the angle averaged absorption probability
in the solar neighborhood. 
The energy dependence of this absorption probability
reflects the spectral shape of the target photon distribution.
The maximum at $E \simeq 2.2$~PeV, and the two shoulders at
150 and 1.6~TeV correspond to interactions with the 
photons of the three main components (CMBR, dust and star emission)
of the target radiation field.
The probability of interactions with the photons of a single component
has a maximum for a gamma ray energy of order
$E_\gamma \cdot \langle \varepsilon \rangle \approx m_e^2$, that corresponds
to the c.m. energy of the photon--photon collisions just above the kinematical threshold,
where the pair production cross section has its maximum value
$\sigma_{\gamma\gamma} \simeq \sigma_{\rm Th}/4$
(where $\sigma_{\rm Th} \simeq 6.65 \times 10^{-25}$~cm$^2$ is the Thomson cross section).
At the maximum, the absorption probability takes the value
$K \approx \sigma_{\rm Th} \; N_\gamma/4$
where $N_\gamma$ is the total number density of target photons that form
the component.

Numerically this corresponds to a minimum interaction lengths
(in the solar neighborhood) of order $\lambda_{\rm abs} = K^{-1} \approx 7$~kpc
at energy $E_\gamma \simeq 2.2$~PeV for absorption by the CMBR,
$\lambda_{\rm abs} \approx 100$ ~kpc at $E_\gamma \simeq 150$~TeV for
absorption by the infrared dust emission,
and $\lambda_{\rm abs} \approx 7$ ~Mpc at $E_\gamma \simeq 1.6$~TeV
for absorption by starlight.

The calculation of the optical depth requires a knowledge of the
target radiation field in the entire volume of the Galaxy,
however, for a qualitative understanding, one can note that the spectra
of the target photons have a similar shape in all points of the
Galaxy. The absorption generated by interactions with the CMBR,
with an absorption length of order
10~kpc, that is of same order of the linear size of the Galaxy is
very important for the propagation of photons in the PeV energy range.
The effects of absorption by dust emitted infrared photons,
with a (space and direction dependent) 
absorption length ten times longer (of order 100~kpc),
are smaller but not entirely negligible.
The effects of absorption on stellar light remain always small.

\section{The local diffuse $\gamma$ ray emission}
\label{sec:qlocal} 
As a first step, in this section we will calculate the local diffuse gamma ray emission, that is the
emission in the vicinity of the solar system.
This calculation requires three elements:
(a) a knowledge of the CR fluxes that are directly observable at the Earth,
(b) a description of the relevant targets (gas and radiation) for CR interactions, and
(c) a model for the interaction cross sections.
The crucial point is that the calculation does not need to
model the space dependence of the CR spectra.

In the following we will discuss separately the two main
(hadronic and leptonic) emission mechanisms.

\subsection{Hadronic emission}
\label{sec:hadronic-emission}
\begin{figure}[hbt]
\begin{center}
\includegraphics[width=8.0cm]{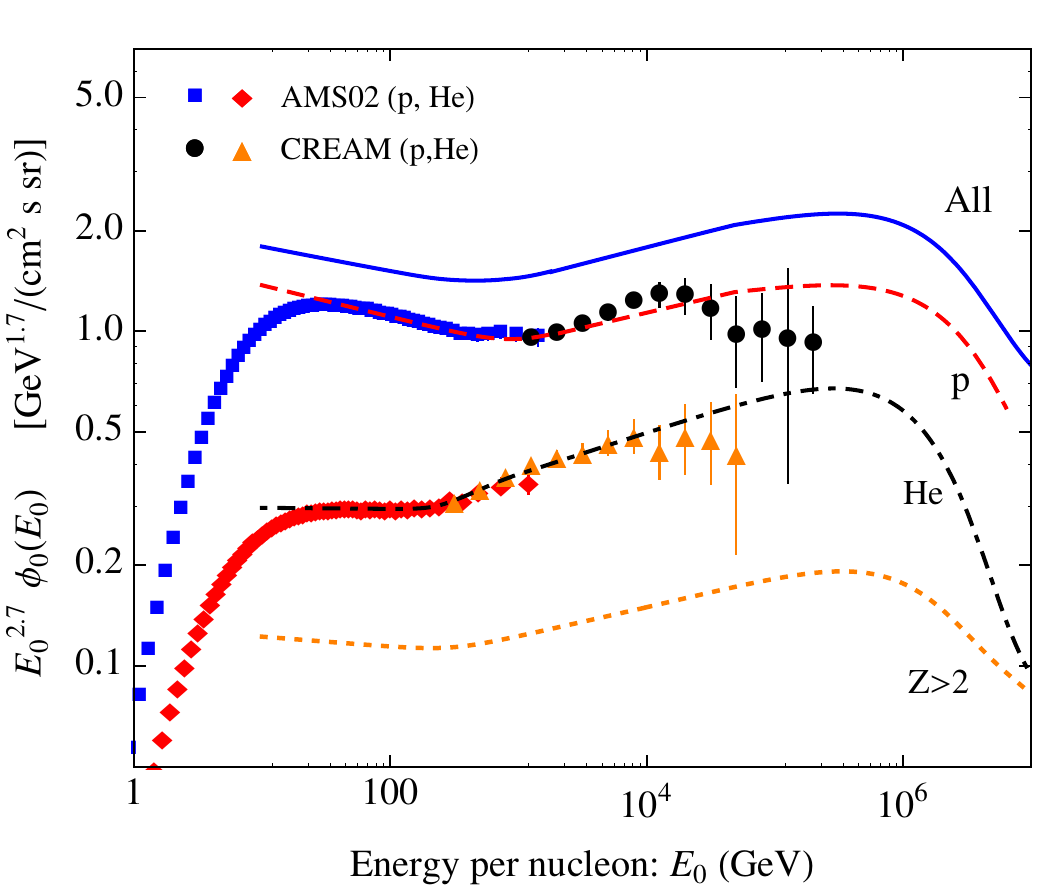}
\end{center}
\caption {\footnotesize
Model of the cosmic ray fluxes at the Earth used in this work.
The lines show the all nucleon flux and the contributions due
to protons, helium, and nuclei with $Z > 2$.
The data points are from
AMS02 \cite{ams02-protons,ams02-helium} and CREAM \cite{Yoon:2017qjx}.
\label{fig:flux-nucleons}}
\end{figure}

The calculation of the hadronic emission requires a description of the
nuclear components of the CR flux.
Fig.~\ref{fig:flux-nucleons} shows our fit to the
observed spectra of protons and nuclei, together with some of the available data.
In the figure the spectra are shown
in the form of the nucleon flux versus the energy per nucleon $E_0$.
The spectra exhibit two evident spectral features. The first one is a
hardening at rigidity $p/Z \approx 300$~GV
that has been observed by CREAM, PAMELA and AMS02
\cite{Ahn:2010gv,pamela-hardening,ams02-protons,ams02-helium}.
The second feature is the well known ``knee''
that is prominent in the all--particle spectrum
observed by air shower experiments for a particle energy $E \simeq 4$~PeV.
Our description of the CR fluxes assumes that also the ``knee'' is
a spectral structure present for all nuclear species
at a constant rigidity, with the softening at 4~PeV corresponding to the break in
the helium component that is dominant at that energy.

For $E_0 \lesssim 10^4$~ GeV our model of the CR spectra
is based on a generalization of the fit given in \cite{Lipari:2017jou},
that gives a good description of the data of AMS02
\cite{ams02-protons,ams02-helium}
and CREAM \cite{Yoon:2017qjx}.
The fluxes of nuclei with $A > 4$ are extrapolated from the
measurement of the HEAO3 detector \cite{Engelmann:1990zz}, introducing a rigidity
dependent hardening.

The CR fluxes observed at the Earth, for $E \lesssim 30$~GeV are distorted by
solar modulation effects. Our model of spectra in interstellar space
are demodulated using the Force Field Approximation with a parameter $V = 0.6$~GeV.
In the present work we discuss only high energy gamma rays, and the description of the
solar modulation effects is of negligible importance.

At high energy, our model of the CR spectra smoothly joins the parametrization of
Gaisser, Stanev and Tilav (GST) \cite{Gaisser:2013bla},
developed to fit the measurements of Extensive Air Shower detectors, and 
the two models become identical for $E_0 > 10^4$~GeV.

The target of the hadronic interactions is the interstellar gas,
and is entirely characterized by its density and chemical composition.
We have assumed the average solar system composition estimated
by Ferri\`ere \cite{Ferriere:2001rg},
with hydrogen, helium and all other nuclei
accounting for fractions 0.90, 0.0875 and 0.0125 of the atoms.
This corresponds to a total mass of the interstellar gas that is 
a factor 1.42 larger than the mass in hydrogen.
The numerical results shown below are calculated for a
nominal value of the hydrogen density $n = 1$~cm$^{-3}$, to allow a simple rescaling. 
In the following (see section~\ref{sec:hydrogen})
we will estimate that the average hydrogen density on the Galactic plane
at radius $r = r_\odot \simeq 8.5$~kpc is $n \simeq 1.48$~cm$^{-3}$
(note that this average quantity is not identical to the gas density
in the vicinity of the solar system that is determined by local effects).

The calculation of the hadronic emission requires a model for
particle production in inelastic hadron--hadron collisions.
Because of the present limitations in the understanding of the strong interaction processes,
a non negligible source of uncertainty cannot be avoided.
The interaction probabilities of relativistic hadrons are 
well understood, because the 
$pp$ total, elastic and inelastic cross section have
been measured with good precision in the entire energy range of interest.
The cross sections for $p$--nucleus and nucleus--nucleus collisions
have been estimated from the data on $pp$ collisions
using a standard Glauber calculation \cite{Glauber-1970}.
Collisions with helium and other nuclei
(nuclei with $A > 4$ have been modeled as entirely formed by oxygen)
account for a weakly energy dependent fraction of order 21--23\% of all
inelastic interactions.

For the description of the inclusive spectra of final state particles,
in the lower energy range ($E_0 < 50$~GeV) we have used the
algorithms described in \cite{Lipari:2016vqk}.
At higher energy we have modeled the interactions using
the Pythia Montecarlo code \cite{Sjostrand:2006za}.

The local gamma ray emission from hadronic interactions is shown in Fig.~\ref{fig:local-emission}.
The spectrum exhibits a strong suppression for $E \lesssim 1$~GeV, 
that has a simple and well known kinematical origin,
associated to the fact that most of the gamma rays are generated
in the decay of neutral pions.
It is straightforward to show that if the photons are entirely generated
by the decay of neutral pions, the $\gamma$ spectrum has
the symmetry: $\phi_\gamma (E_\gamma)= \phi_\gamma (m_{\pi}^2/(4 \, E_\gamma))$
(with $m_\pi$ the $\pi^\circ$ mass).
This property implies that the spectrum has a maximum at $E = m_{\pi}/2$,
and that the emission of photons with energy $E \lesssim m_\pi$
is strongly suppressed. The observation of this spectral
feature can in principle be used to identify in a model independent
way the hadronic component in the gamma ray flux.

At higher energy ($E_\gamma \gg m_{\pi}$) the hadronic gamma ray emission
falls roughly as a power law, but the spectral index is not
constant and changes gradually with energy.
Inspecting Fig.~\ref{fig:local-emission} one can note
a gradual hardening of the spectrum centered at $E_\gamma \approx 50$~GeV,
and then a more marked, but also gradual softening
centered at $E_\gamma \approx 2 \times 10^5$~GeV.
These broad structures reflect
the features present in the primary CR spectra
that we have discussed above.

For a qualitative understanding one can note that
for $E \gg 1$~GeV, the inclusive spectra of the final state particles,
created in inelastic hadronic collisions,
have (in reasonably good approximation) the scaling property:
\begin{equation}
\frac{dN_{pp \to \gamma}}{dE} (E, E_0) \simeq \frac{1}{E_0} ~F_{pp \to \gamma}
 \left ( \frac{E}{E_0} \right ) ~.
\end{equation}
This equation can be derived assuming the validity 
of Feynman scaling in the projectile fragmentation region, and
has the well known consequence that if the
spectrum of primary nucleons is a simple power law
with exponent $\alpha$, the emission is
then a power law with the same spectral index.
The violations of Feynman scaling, and the growth of the
inelastic cross sections with c.m. energy introduce logarithmic corrections.

The all nucleon flux cannot however be
described as a simple, unbroken power law because of the existence of
the spectral features that we have discussed above:
the ``Pamela hardening'' and the ``knee''.
The existence of these structures in the primary CR spectra 
are reflected in more gradual features in the $\gamma$ emission, that are
centered at (a factor of order 5--10) lower energy.
This can be easily understood noting 
that photons with energy $E_\gamma$ are generated by
the interactions of primary nucleons with in a broad range of energy,
with an order of magnitude extension and 
a median value $E_0 \sim 6 ~E_\gamma$ (the precise value
depends on the spectral index).

\subsection{Leptonic emission}
\begin{figure}[hbt]
\begin{center}
\includegraphics[width=8.0cm]{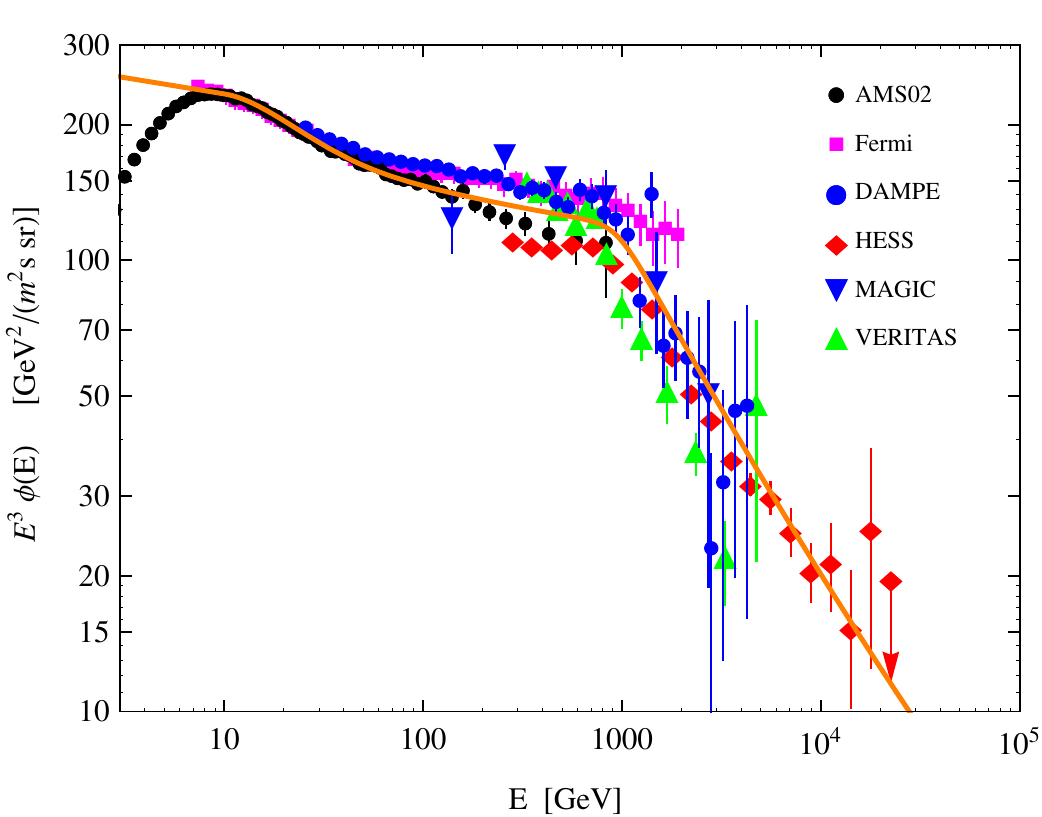}
\end{center}
\caption {\footnotesize
Flux of electrons plus positrons observed at the Earth.
The line is our fit to the spectrum.
The data points are from
Fermi \cite{Abdollahi:2017nat},
AMS02 \cite{ams-all-electrons},
DAMPE \cite{Ambrosi:2017wek},
HESS \cite{Aharonian:2008aa,Aharonian:2009ah,hess-icrc2017},
MAGIC \cite{BorlaTridon:2011dk} and
VERITAS \cite{Staszak:2015kza}.
\label{fig:flux-electrons}} 
\end{figure}

The calculation of the leptonic emission requires a description
of the flux of electrons plus positrons.
Our fit to the ($e^- + e^+)$ is shown
in Fig.~\ref{fig:flux-electrons} together with some of the measurements.
The flux is accurately measured for $E_e \lesssim 500$~GeV
by the observations of detectors on satellites
like PAMELA \cite{pamela-electrons,pamela-positrons,Adriani:2014pza}
Fermi \cite{Abdollahi:2017nat} and AMS02 \cite{ams-all-electrons}.
The observations of HESS \cite{Aharonian:2008aa,Aharonian:2009ah,hess-icrc2017},
and later by MAGIC \cite{BorlaTridon:2011dk}, VERITAS \cite{Staszak:2015kza},
and more recently by DAMPE \cite{Ambrosi:2017wek} have shown that the spectrum
has a break at $E \approx 900$~GeV where the spectrum steepens
from a spectral index of order 3.1 to an index of order 3.8.

Two mechanisms contribute to the leptonic emission.
In bremsstrahlung the target (interstellar gas) is identical
to the one discussed for the hadronic emission.
For Compton scattering the target are the photons of the
interstellar radiation fields discussed in sec.~\ref{sec:absorption}.
For our calculations we have used the model of \cite{Vernetto:2016alq}.

The leptonic mechanisms for gamma ray production
are purely electromagnetic and therefore have exactly calculable cross sections
(see for example \cite{Blumenthal:1970gc}).

\begin{figure}[hbt]
\begin{center}
\includegraphics[width=8.0cm]{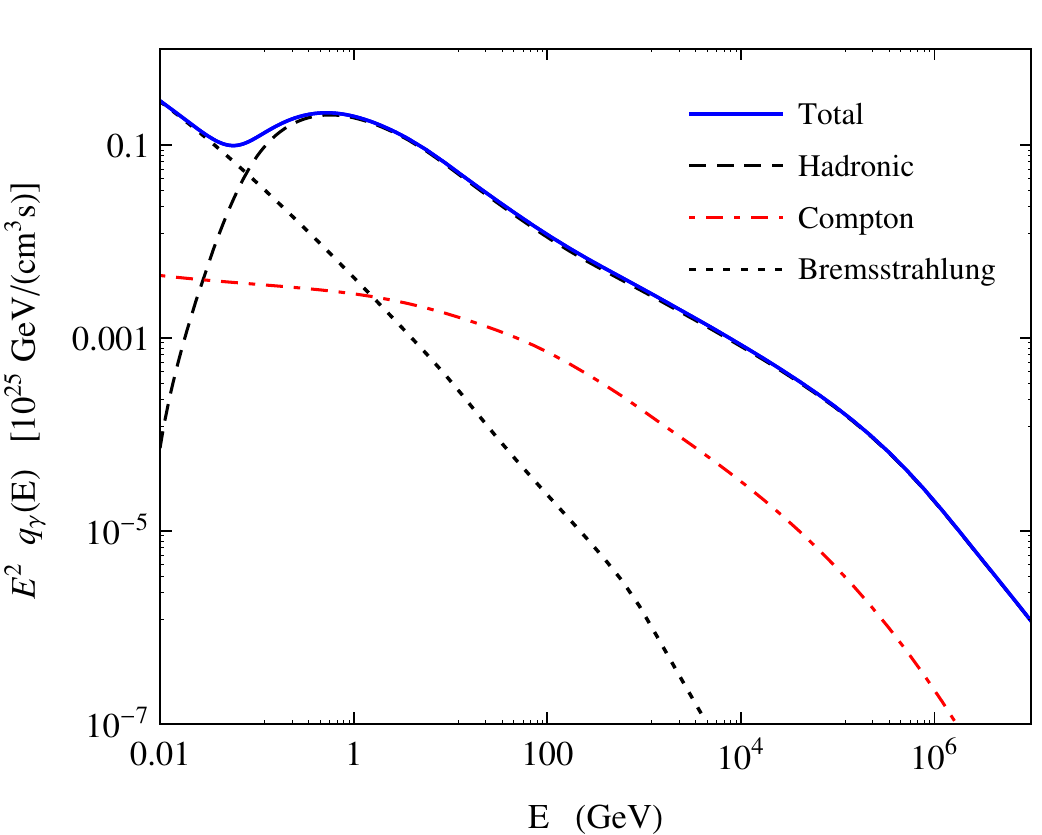}
\end{center}
\caption {\footnotesize
Gamma ray emission rate in the solar neighborhood.
 The total emission and the single contributions of the three main
 mechanisms (hadronic, bremsstrahlung and Compton scattering) are shown. 
The assumed hydrogen density in the interstellar medium is $n = 1$~cm$^{-3}$.
 \label{fig:local-emission}}
\end{figure}

The leptonic emission, separated into the contributions of bremsstrahlung and
Compton scattering, is shown in Fig.~\ref{fig:local-emission}.
The results can be easily understood qualitatively.
In the case of bremsstrahlung, the $e^\mp$ radiates photons
with an energy independent cross section and the final state photon
has an energy distribution
that depends only on the ratio $E_\gamma/E_e$.
If the primary $e^\mp$ have a power law spectrum, the emission is then
also a power law with the same spectral index
(of order $\alpha_e \approx 3.1$ for $E \lesssim 300$~GeV).
The bremsstrahlung spectrum softens at higher energy
because of the break in the ($e^- + e^+)$ spectrum at $E_e \approx 1$~TeV.

The Compton scattering component of the emission has initially
a hard spectrum (a spectral index of order 2).
This reflects the well known fact that when the $e\gamma$ interactions are
in the Thomson regime (that is when the product
of the energies of the interacting particles is sufficiently small:
$E_e \; \varepsilon_i \lesssim m_e^2$ ) the spectral indices of the Compton
emission and the primary electron flux are related by:
$\alpha_\gamma \simeq (\alpha_e +1)/2$.
This behaviour however stops for $E_\gamma \gtrsim 100$~GeV
when most of the $e\gamma$ interactions are in
the Klein--Nishina regime. The Compton emission suffers more
suppression at higher energy also 
because of the softening of the $e^\mp$ spectra above 1~TeV.
The result is that the local Compton emission of gamma rays
gives a maximum contribution of order 5\%
with respect to the hadronic one.

It should be noted that the estimate
of the contribution of the leptonic mechanisms to the observed
gamma ray flux (that is the result of the emission from the whole Galaxy)
is a more difficult task,
because it requires to compute the emissions in different regions
of the Galaxy, where the densities of the primary particles (electrons, protons
and nuclei) and of the relevant targets (gas and radiation) can be different.
In particular it is possible, and indeed likely
that the Compton mechanism can be a significant
component of the flux for directions that point away from the
Galactic equator. This is because the interstellar gas density
(the target for hadronic emission) is exponentially
suppressed for large $|z|$, while the density of the radiation fields
(the target for Compton scattering)
falls more gradually with $|z|$ (and the CMBR component is in fact
constant). However one expects that
in the region of small latitude
($|b| \lesssim 10^\circ$), where the diffuse flux is largest,
the leptonic mechanisms remains subdominant, with a
maximum contribution of order $\lesssim 10\%$ to the diffuse flux.

\subsection{Summary}

The main results obtained in this section can be summarized as follows:
\begin{itemize}
\item[(A)] The hadronic mechanism is the dominant source
 of gamma rays for $E_\gamma \gtrsim $~few~GeV in the Galactic disk region
 ($|b| \lesssim 10^\circ$) with the leptonic mechanisms accounting
 for only a few percent of the emitted photons.
\item[(B)] In a broad energy interval
 between 10 and $10^5$~GeV the emission spectrum
 is in good approximation a power law ($q_\gamma \propto E^{-\alpha}$)
 with a spectral index $\alpha$ that varies slowly with values between 2.8 and 2.6.
\item[(C)] For $E_\gamma$ between $10^5$ and $10^6$~GeV
 the energy distribution of the emission softens,
 and the spectral index grows gradually up a value of order 3.1.
 Between 1 and 10~PeV the spectral index remains approximately constant.
\end{itemize}

\section{Galactic interstellar gas density}
\label{sec:hydrogen}

To model the diffuse gamma ray emission in the whole Galaxy, a
knowledge of the spatial distribution of the density and
composition of the interstellar medium is required. 
The interstellar matter has been 
the object of many studies, and several large scale surveys of its main components
have been performed in recent years.
These studies have revealed that the space distributions of interstellar gas and dust
have a complex form, with structures present at many scales.

Our goal in the present work is to construct a model of the interstellar
gas density that captures reasonably well its large scale properties,
without attempting to
describes accurately finer details such as individual clouds and filaments.
For this purpose we have assumed an axially and up-down symmetric distribution
of the gas that neglects the spiral arms, whose geometry remain controversial,
and north-south asymmetries such as the disk warp.

The model is constructed using previous studies on the
distribution of neutral atomic (H), molecular (H$_2$) and ionized hydrogen, 
that are the most important components of the interstellar medium.
The contribution of other elements is added to the hydrogen
assuming that the composition of the interstellar gas is equal
in the entire Galaxy, and equal to the composition
of the solar system estimated by
Ferri\`ere \cite{Ferriere:2001rg} (see section~\ref{sec:hadronic-emission}).

The interstellar gas is confined to a narrow region around the
Galactic plane with a thickness that grows with $r$ (the so called ``disk flaring'').
Assuming for the $z$ dependence a Gaussian form, the hydrogen
density can be written as:
\begin{equation}
n(r,z) = n_0 (r) ~ \exp \left [ - \frac{z^2} {2 \, \sigma_z^2(r)} \right ]
\end{equation}
with $r$ and $z$ cylindrical coordinates.
In this equation $n_0(r)$ is the midplane density, and
$\sigma_z (r)$ is related to the half width at half maximum (HWHM) of the
$z$ distribution $H_W(r)$ by the relation $H_W = \sqrt{2 \, \ln 2} ~\sigma_z$.
An important quantity is the surface density 
\begin{equation}
 \Sigma(r) = \int_{-\infty}^{+\infty} dz~n(r,z)
\end{equation}
that is related to the midplane density an the disk thickness by the relations:
\begin{equation}
\Sigma (r) = \sqrt{2 \, \pi} \, n_0(r) \, \sigma_z (r)
= \sqrt{\pi/\ln 2} \; n_0 (r) \, H_W (r)
\end{equation}

\subsection{Molecular hydrogen}
Molecular hydrogen cannot be observed directly, but its spatial distribution can be 
inferred through the observation of the emission lines of carbon monoxide (CO).

To describe the $H_2$ distribution in the Galactic disk, excluding the central region, we use the results of Roman-Duval et al. \cite{RomanDuval:2016}, that
report the radial distributions of the surface density and the disk
thickness for $r$ from 2.5 to 16~kpc, obtained analyzing $^{12}$CO and $^{13}$CO data.
Their evaluations are consistent with the
measurements reported in the review by Heyer and Dame \cite{Heyer:2015}.
According to their results, the H$_2$ surface density has a maximum 
at $r = 4.25$~kpc. Smoothing the small scale granularity of the observed distribution, the radial dependence of the surface density can be approximately
described by exponential functions, with a slope change at $r = 4.25$~kpc.
According to the data reported in the same paper,
the disk thickness for $r < 8$~kpc is approximately constant,
$H_W(r) \simeq 50$~pc, and grows exponentially at larger radii.

Concerning the central region of the Galaxy, we use the work of Ferriere et al.
\cite{Ferriere:2007}, who model the molecular gas distribution 
by combining different sets of data.
According to their study, in the Galaxy Center the most prominent feature is a small region of radius 
$\sim$200 pc with a very high H$_2$ density, the Central Molecular Zone (CMZ), 
actually centered at $\sim$50 pc from the Galactic Center,
embedded in a lower density ring extending up to $\sim 1.5$~kpc.
Both these structures have an elongated shape 
and are tilted with respect to the Galactic plane and to the line of sight. 
Since their precise geometry is not known
and since we only need to describe the
general shape of the gas distribution, for simplicity we model the central region
assuming an axisymmetric behaviour, with the surface density exponentially
decreasing with $r$. To find the radial slope and normalization, we set
the total H$_2$ mass in the (CMZ + ring) region equal to the mass 
evaluated by Ferriere et al. (5.3 10$^7$ M$_{\odot}$) 
and the surface density at $r = 1.5$ kpc equal to the
value obtained by the function used for the disk, previously described.

Concerning the distribution along $z$, according to Ferriere et al.
the thicknesses of the CMZ and the ring are approximately costant, 
but the CMZ is thinner ($H_W =15$~pc) than the ring ($H_W=35$~pc).
We set $H_W = 15$~pc at the center, and assume a flaring
in order to connect smoothly with the data at larger radii.

Our parametrizations of the surface density, midplane density $n_0(r)$
and disk thickness $H_W(r)$ for molecular hydrogen
are listed in Table~\ref{tab:tabH1}.
These parametrizations have a discontinuous derivative for some values of $r$.
In our numerical calculations
the discontinuities are smoothed with a length scale $\Delta x = 50$~pc.
The model corresponds to a total mass of molecular hydrogen of $7.2 \cdot 10^8$~M$_{\odot}$.

\begin{table}[ht!]
 \begin{center}
 \begin{tabular}{| l | l | l | l |} 
 \hline
 $r$ [kpc] & $\Sigma(r)$ [$M_\odot$ pc$^{-2}$] & $n_0(r)$ [cm$^{-3}$] & $H_W(r)$ [pc]\\
 \hline
 0 - 1.5 & 105 $e^{-r/0.29}$ & 135 $e^{-r/0.235}$ & 15 $e^{r/1.25}$\\
 1.5 - 4.25 & 0.598 $e^{(r-1.5)/1.09}$ & 0.229 $e^{(r-1.5)/1.09}$ & 50 \\
 4.25 - 8 & 7.5 $e^{-(r-4.25)/1.86}$ & 2.88 $e^{-(r-4.25)/1.86}$ & 50 \\
 $>$ 8 & 1.0 $e^{-(r-8)/1.56}$ & 0.383 $e^{-(r-8)/1.10}$ & 50 $e^{(r-8)/3.7}$\\
 \hline
 \end{tabular}
 \caption{H$_2$ surface density, midplane density and vertical thickness (HWHM) in different radial regions.
 The radius $r$ is in kpc.
 \label{tab:tabH1}
}
 \end{center}
\end{table}

\subsection{Atomic hydrogen}
The distribution of neutral atomic hydrogen is studied throught the
21-cm radio line, emitted in the transition between the two hyperfine levels
of the $1s$ ground state.
We model the atomic hydrogen distribution according to the measurements 
reported by Kalberla and Dedes \cite {Kalberla:2008uu} for the region outside
the Galactic Bulge, who fit the midplane density for r $>$ 8 kpc with the exponential form:
\begin{equation}
n_0 (r) = n_\odot ~e^{-(r- r_\odot)/R}
\end{equation}
where $R = 3.15$~kpc and $n_\odot = 0.9$~cm$^{-3}$ is the density at the Sun radius.

The midplane density is approximately constant between 4 and 8~kpc, and 
decreases rapidly towards the center of the Galaxy.

The vertical scale of the atomic hydrogen distribution $H_W(r)$ grows exponentially with $r$.
Kalberla and Dedes describe the radial dependence of $H_W(r)$ as 
\begin{equation}
 H_W (r) = h_0 ~e^{-(r-r_\odot)/R_h}
\label{eq:hr2}
\end{equation}
and fit the observations with $h_0 = 0.14$~kpc and $R_h = 9.8$~kpc.

The density of atomic hydrogen in the central zone of our Galaxy is
about one order of magnitude lower than the molecular component.
Similarly to the molecular case, we derive the H distribution starting from
the evaluations by Ferriere et al.
We describe the surface density for $r < 1.5$~kpc with an exponential
function having the same slope obtained for the molecular gas. We fixed the normalization
in order to have the total mass of atomic hydrogen as reported in the same paper
(5.2 10$^6$ M$_{\odot}$). We connect the density
at $r$=1.5 kpc to the density at $r$=4 kpc (provided by the Kalberla and Dedes fit)
with a further exponential curve.

Concerning the gas thickness, we set $H_W = 45$~pc at the Galaxy Center 
(the value given by Ferriere et al. for the CMZ) and assume a flaring in order
to connect smootly with the curve of equation~(\ref{eq:hr2}).

Table~\ref{tab:tabH2} summarizes our
parametrizations for atomic hydrogen in the Galaxy.
As in the case of molecular hydrogen, discontinuities in the derivative of
the functions are smoothed in the numerical calculations. 
The corresponding total mass of atomic hydrogen is $5.3 \cdot 10^9$~M$_{\odot}$.

\begin{table}[ht!]
 \begin{center}
 \begin{tabular}{| l | l | l | l |} 
 \hline
 $r$ [kpc] & $\Sigma(r)$ [$M_\odot$ pc$^{-2}$] & $n_0(r)$ [cm$^{-3}$] & $H_W(r)$ [pc]\\
 \hline
 0 - 1.5 & 10.2 $e^{-r/0.29}$ & 3.21 $e^{-r/0.265}$ & 45 $e^{r/3.06}$\\
 1.5 - 4 & 0.058 $e^{(r-1.5)/0.52}$ & 0.011 $e^{(r-1.5)/0.549}$ & 150 $e^{(r-8.5)/9.8}$ \\

 4 - 8 & 7.07$e^{(r-4.)/9.8}$ & 1.05 & 150 $e^{(r-8.5)/9.8}$ \\
 $>$ 8 & 10.6 $e^{-(r-8)/4.64}$ & 1.05 $e^{-(r-8)/3.15}$ & 150 $e^{(r-8.5)/9.8}$ \\
 \hline
 \end{tabular}
 \caption{
 \label{tab:tabH2}
Atomic H surface density, midplane density and vertical thickness (HWHM)
in different radial regions.
The radius $r$ is in kpc.}
 \end{center}
\end{table}

\subsection{Ionized hydrogen}
The density of ionized hydrogen has beeen modeled by \cite{Taylor:1993my,Cordes:2002wz}.
In most of the Galaxy this component of the interstellar gas is negligible,
however in the vicinity of the Galactic Center it is comparable to the
contribution of neutral atomic hydrogen.

\vspace{0.5 cm}

\begin{figure}[hbt]
\begin{center}
\includegraphics[width=8.0cm]{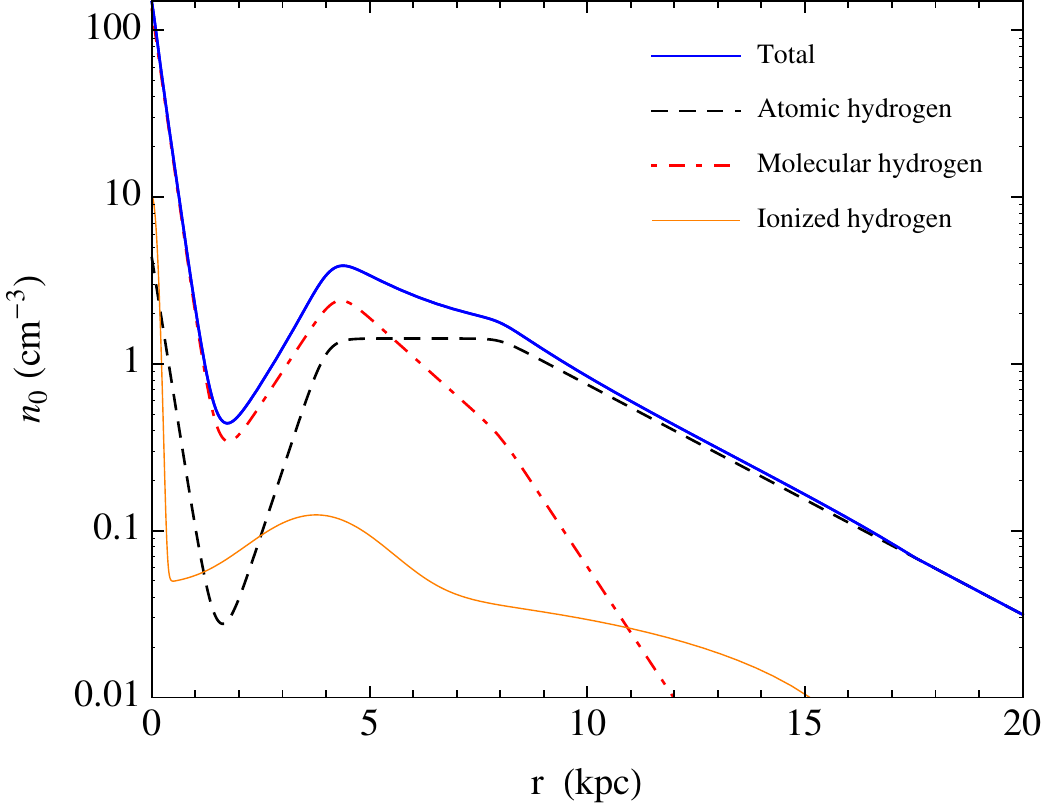}
\end{center}
\caption {\footnotesize
Midplane density of hydrogen gas $n_0(r)$.
The single contributions of atomic, molecular and ionized hydrogen are
also shown. 
 \label{fig:dens-midplane}}
\end{figure}

Fig.\ref{fig:dens-midplane} shows the midplane density of atomic, molecular and total hydrogen
as a function of $r$, for the whole Galactic plane.

\section{Fermi observations of the diffuse Galactic gamma ray flux} 
\label{sec:fermi} 

As a starting point for a model of the gamma ray diffuse emission 
at TeV--PeV energies, we use the existing data in the GeV energy range.

The latest and most accurate measurements of the diffuse flux in the
energy range 0.1--100~GeV have been obtained in the last few years
by the Fermi telescope. The Fermi collaboration has published in 2012 
a dedicated paper about the diffuse Galactic emission
\cite{Ackermann:2012pya} , however a significant amount of
data has been obtained after that,
and the methods of analysis have also significantly improved. 
Some of these new results are discussed in \cite{Acero:2016qlg}.

The Fermi data are public, and several authors
(for example \cite{Gaggero:2014xla,Yang:2016jda})
have performed independent studies of the Galactic diffuse flux.
In the present work we will not perform an independent analysis of the Fermi data,
to estimate the diffuse gamma ray flux.
This is a very important but difficult task that is postponed to a future work.

The Fermi Collaboration has made available
a template of the diffuse Galactic gamma ray flux
to be used as a background model for the search of
point sources \cite{fermi-background}.
This background model gives tables of the angular distribution of the flux
(in bins equispaced in Galactic
latitude and longitude with a linear size 0.125$^\circ$) for a discrete set of
30 energies (equispaced in $\log E$) between 58.5 MeV and 513 GeV.

In this paper we will use the Fermi background model
as a first order approximation of the Galactic diffuse flux.
We have chosen the map at the energy $E^* = 12$~GeV (more precisely 11.98~GeV)
as a template of the angular distribution of the real diffuse flux at the same energy.
This template will be used here as a ``boundary condition'' for
extrapolations to higher energies.
The reference energy $E^*$ has been chosen as a reasonable
optimum choice on the basis of the following considerations. \\
(i) The energy must be sufficiently high, so that the contributions of
the leptonic mechanisms to the gamma ray flux is small
(see discussion in sec.~\ref{sec:qlocal}). \\
(ii) The energy must be sufficiently low, so that the diffuse
flux is measured with good statistical accuracy.
A low value of $E^*$ is also desirable because it
allows to study the evolution of the diffuse flux in a broader energy range,
when constructing different models for the extrapolation to very high energy.

\section{Model 1: space independent CR spectra} 
\label{sec:factorized}

In the most commonly accepted models for Galactic cosmic rays, 
the spectral shape of the nuclear components 
(protons and nuclei) are identical in the entire volume of the Milky Way.
This result emerges in a large class of models where the following conditions are satisfied:
\begin{enumerate}
\item The populations of CR in the Galaxy are in a stationary
state, with the sources that compensate the losses due to escape and
other effects. The spectra are not
significantly distorted by the contributions of near sources that are still active
or have been active in the recent past
(time variations of the CR spectra
associated to the evolution of the Galaxy can exist for cosmological time scales).
\item The spectra generated by the CR sources in different regions of the
 Galaxy have, after an appropriate average in time, a space independent shape.
 This condition is immediately satisfied in models where
 a single class of astrophysical events
(for example SN explosions or GRB's) is the dominant CR source.
\item CR propagation is well described by
diffusion with a diffusion coefficient that has the same
rigidity dependence in all points of the Galaxy.
This corresponds to the statement that the space
and rigidity dependences of the diffusion coefficient 
are factorized: $D(p/q, \vec{x}) \simeq D_0(p/q) \; f_D(\vec{x})$.
\item Escape from the Galaxy is the dominant mechanism for CR losses. 
\item Energy losses during CR propagation are negligible.
\end{enumerate}
It is straightforward to see that if these conditions
are satisfied the CR spectral shapes are proportional to the
ratio of the time averaged source spectrum and the
rigidity dependence of the diffusion coefficient.
For example, for ultrarelativistic protons
$\phi_p (E, \vec{x}) \propto Q_p (E)/D_0(E)$.
The absolute normalization of the CR fluxes
is in general a function of position $\vec{x}$, and 
depends on the geometry of the Galactic confinement volume and on
the space distribution of the sources.
Several authors have published interpretations of the CR data 
based on these assumptions, estimating the spectral shape of the source and
the rigidity dependence of the diffusion coefficient.

The hypothesis that the nucleon flux has a spectral shape
that is independent from the position can be written in the form:
\begin{equation}
 \phi_0 (E_0, \vec{x}) = \phi_0^{\rm loc} (E_0) \times f_0 (\vec{x})
\label{eq:phi-factorization}
\end{equation}
where $\phi_0^{\rm loc} (E_0)$ is the locally observed spectrum,
and $f_0 (\vec{x})$ an adimensional function of the space coordinates
that, by construction, satisfies the constraint $f_0 (\vec{x}_\odot) = 1$.
Eq.~(\ref{eq:phi-factorization}) implies that
the emission of gamma rays (and neutrinos) generated by the
hadronic mechanism has also a factorized form:
\begin{equation}
 q_\gamma (E,\vec{x}) =
 q_\gamma^{\rm loc} (E) \times f_0(\vec{x}) \times
 \left ( \frac{n(\vec{x})}{n(\vec{x}_\odot)} \right )
\label{eq:q-factorized}
\end{equation}

Inserting this factorized form of the emission in the general
expression for the diffuse gamma ray flux 
of Eq.~(\ref{eq:phi-diffuse}), and 
assuming that the absorption effects are negligible (that is in the limit
$\tau \to 0$) one finds that the 
energy and angular dependences of the $\gamma$ ray flux are factorized:
\begin{equation}
 \phi_\gamma (E, \Omega) = \frac{q_\gamma^{\rm loc} (E)}{4 \, \pi} ~T(\Omega) ~
\label{eq:factor0}
\end{equation}
where $T(\Omega)$ is a direction dependent length,
given by the integral:
\begin{equation}
 T(\Omega) = \frac{1}{n(\vec{x_\odot})} ~
 \int_0^\infty dt~ f_0 (\vec{x}_\odot + t \; \hat{\Omega}) \times
 n(\vec{x}_\odot + t \; \hat{\Omega}) ~.
\end{equation}

The inclusion of absorption effects
introduces an energy dependent distortion of the
angular distribution of the flux, and therefore breaks
the factorization of Eq.~(\ref{eq:factor0}).
A more general expression for the gamma ray flux can be written in
the form:
\begin{equation}
 \phi_\gamma (E, \Omega) = \frac{q_\gamma^{\rm loc} (E)}{4 \, \pi} ~T(\Omega)
 \left \langle P_{\rm surv} (E, \Omega) \right \rangle 
\label{eq:factor1}
\end{equation}
where we have indicated with $\langle P_{\rm surv} (E, \Omega) \rangle$ the
gamma ray survival probability, averaged over all points along the line of sight.

The crucial point of this discussion is that if the CR spectra
have a space independent spectral shape, then the diffuse gamma ray flux
has dependences on the energy and angle that are factorized
when absorption effects are negligible.
The factorization is broken at high energy
($E \gtrsim 30$~TeV) when the absorption probability becomes significant.

Starting from these assumptions, we will evaluate the absolute gamma ray flux 
$\phi_\gamma (E, \Omega)$, using the interstellar gas model discussed 
in section~\ref{sec:hydrogen}, and introducing a simple parametrization for
the space dependence of the CR flux:

\begin{equation}
 f_0 (\vec{x}) =
 f_0 (r,z) = \frac{{\rm sech}(r/R_{\rm cr}) ~{\rm sech} (z/Z_{\rm cr})}{{\rm sech}(r_\odot/R_{\rm cr})}
~.
 \label{eq:space-cr}
\end{equation}
In this expression the function ${\rm sech}\, x = (\cosh x)^{-1}$ is the hyperbolic secant.
This function falls exponentially for large values of the argument, but its
derivative vanishes at $x =0$ as it is expected for the CR density at $r= 0$ and $z=0$. 

In the framework we are discussing in this section,
if the density of interstellar gas is known,
the calculation of the gamma ray flux is a straightforward exercise,
with results that are entirely determined by the
two parameters $R_{\rm cr}$ and $Z_{\rm cr}$ associated to
the space dependence of the CR fluxes.

\begin{figure}[hbt]
\begin{center}
\includegraphics[width=8.0cm]{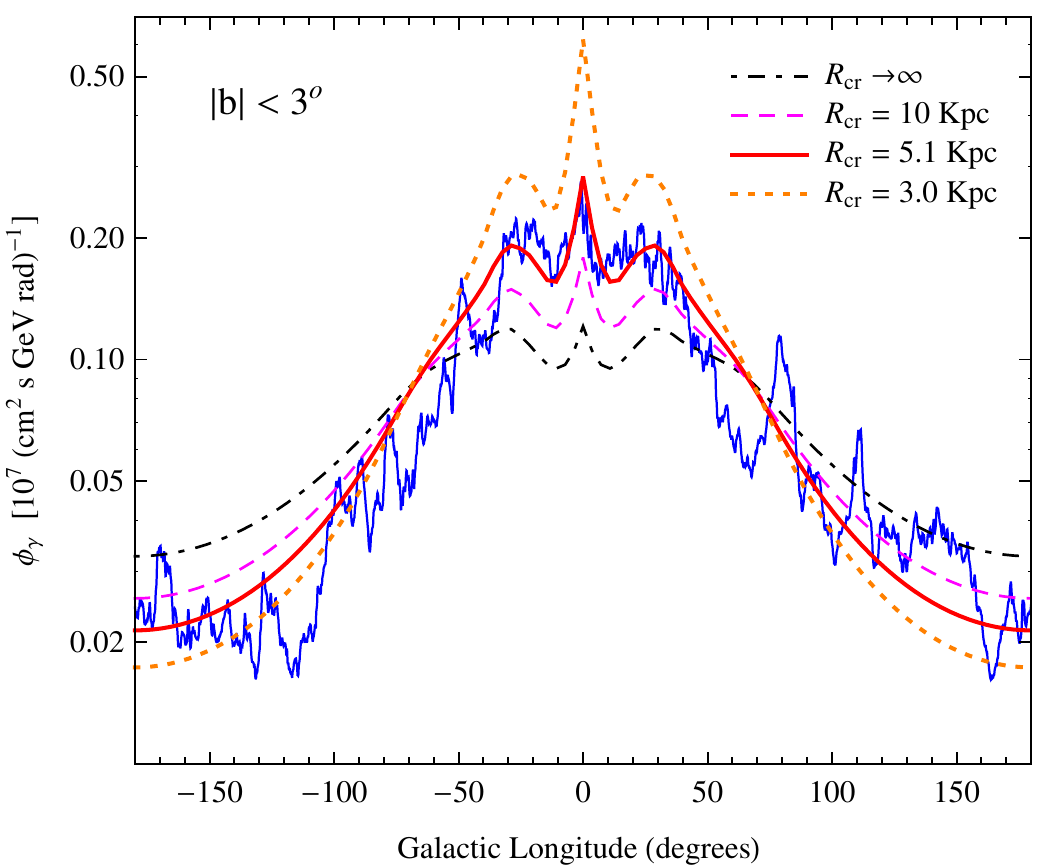}
\end{center}
\caption {\footnotesize
 Longitude distribution for the gamma ray flux
 at $E^* = 12$~GeV integrated in the latitude range $|b| < 3^\circ$.
The model is calculated for different values of the parameter $R_{\rm cr}$.
The calculation is compared to the Fermi background template.
 \label{fig:llong1}}
\end{figure}

One example of the angular distribution of the gamma ray flux for the reference energy $E^*$
calculated under the factorization hypothesis for the CR spectra, and using the
model of interstellar gas discussed in section~\ref{sec:hydrogen}, 
is shown in Fig.~\ref{fig:llong1} and compared to the Fermi template.
The figure shows the longitude distribution of the flux 
after integration in the latitude range $|b| < 3^\circ$, for different values of $R_{\rm cr}$.
Since the gas target is confined in a narrow region in $|z|$, the calculation
is insensitive to the value of $Z_{\rm cr}$ if the parameter is sufficienty large
($Z_{\rm cr} \gtrsim 0.3$~kpc).

Inspecting Fig.~\ref{fig:llong1} one can see that the
Fermi template for the diffuse gamma ray flux
has a rich structure with significant variations for angular scales as small as one degree.
These rapid variations of the flux are consistent with the hypothesis that 
the gamma ray emission is proportional to the density of a very clumpy interstellar gas.
Our calculation cannot reproduce the detailed structure of the
flux for small angular scales, however it can describe
reasonably well the large scale structure of the flux.

The hypothesis that the CR density is constant in the Galactic disk
(and equal to what is observed at the Earth) can be excluded
because such a model, that corresponds to the limit $R_{\rm cr} \to \infty$,
predicts a flux that it too small for directions toward the Galactic Center, and
too large for directions in the opposite hemisphere.
A finite $R_{\rm cr}$, that corresponds to a space gradient of the CR flux 
with a larger density in the GC region, 
results in a better agreement of our model with the Fermi template.
The value $R_{\rm cr} \simeq 5.1$~kpc gives approximately the correct
ratio for the contributions of the two hemipheres toward the Galactic
Center and Anticenter.

\begin{figure}[hbt]
\begin{center}
\includegraphics[width=8.0cm]{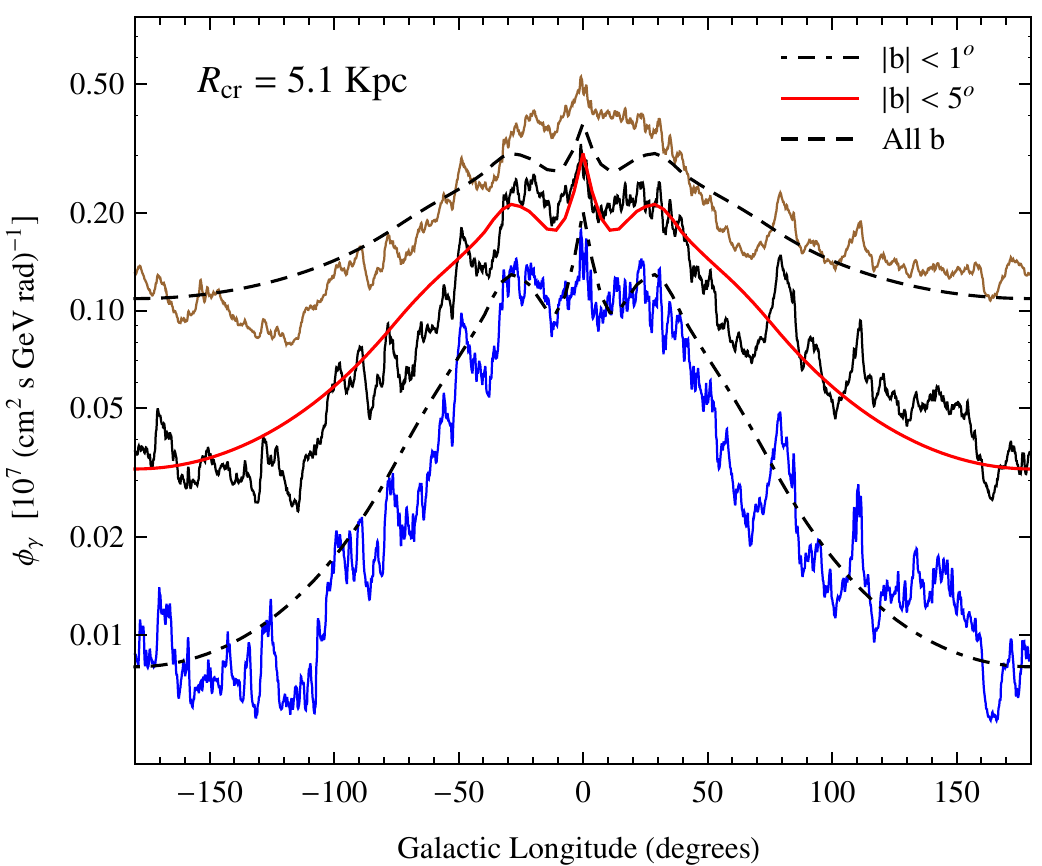}
\end{center}
\caption {\footnotesize
Longitude distribution of the gamma ray flux
at $E^*=12$~GeV calculated in our model and compared
to the Fermi background template.
The calculation uses the parameter $R_{\rm cr} = 5.1$~kpc
to describe the space dependence of the CR density. 
The flux is integrated in three different ranges of latitude.
\label{fig:llong2}}
\end{figure}

\begin{figure}[hbt]
\begin{center}
\includegraphics[width=8.0cm]{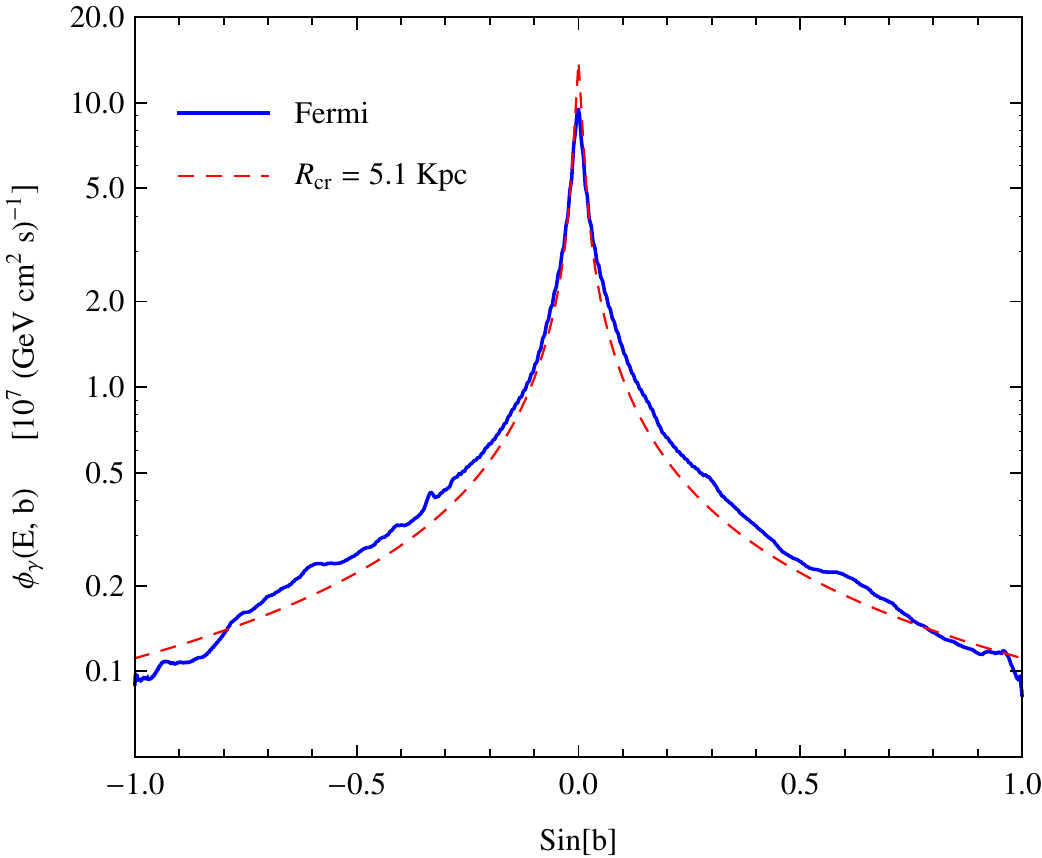}
\end{center}
\caption {\footnotesize
 Latitude distribution of the gamma ray flux
 at $E^* = 12$~GeV according to our model,
 compared to the Fermi background template.
 \label{fig:llat}}
\end{figure}

Other illustrations of the calculated gamma ray flux at the reference energy $E^*$
are given in Fig.~\ref{fig:llong2} that shows the Galactic longitude distribution after integration
over the latitude ranges
$|b| < 1^\circ$,
$|b| < 5^\circ$ and $|b| < 90^\circ$ (that is the entire sky).
Fig.~\ref{fig:llat} shows the latitude distribution of the flux at the reference energy $E^*$,
after integration over all longitudes.

The comparison of the model with the Fermi template shows
that the main features of the diffuse gamma ray flux can be described reasonably well. 
The largest discrepancies are observed at large
$|b|$, and the effect is likely associated to the existence
to the structures known as the ``Fermi bubbles''
\cite{Su:2010qj,Fermi-LAT:2014sfa}.
It should be stressed that the prediction that we have constructed is absolute,
and in fact it is remarkable that a very simple model such as the one we have constructed
can reproduce the observations (in shape and absolute normalization)
with an error of order 10--20\%.

The reasonably good agreement between the model and the Fermi template
for the energy $E^*$ gives support to the idea that hadronic mechanism is the main
source of the diffuse gamma ray flux, and also indicates that the description
of the interstellar gas density is adequate, however it is not sufficient
to conclude that the factorization hypothesis for the CR spectra is correct.
This is because the gamma ray flux at a single energy $E_\gamma$
is sensitive to the spectrum of primary nucleons in a rather narrow
range of energy ($E_0 \approx 3$--30~$E_\gamma$), and in our calculation the
absolute normalization of the CR spectra can have a non trivial space dependence.

\begin{figure}[hbt]
\begin{center}
\includegraphics[width=8.0cm]{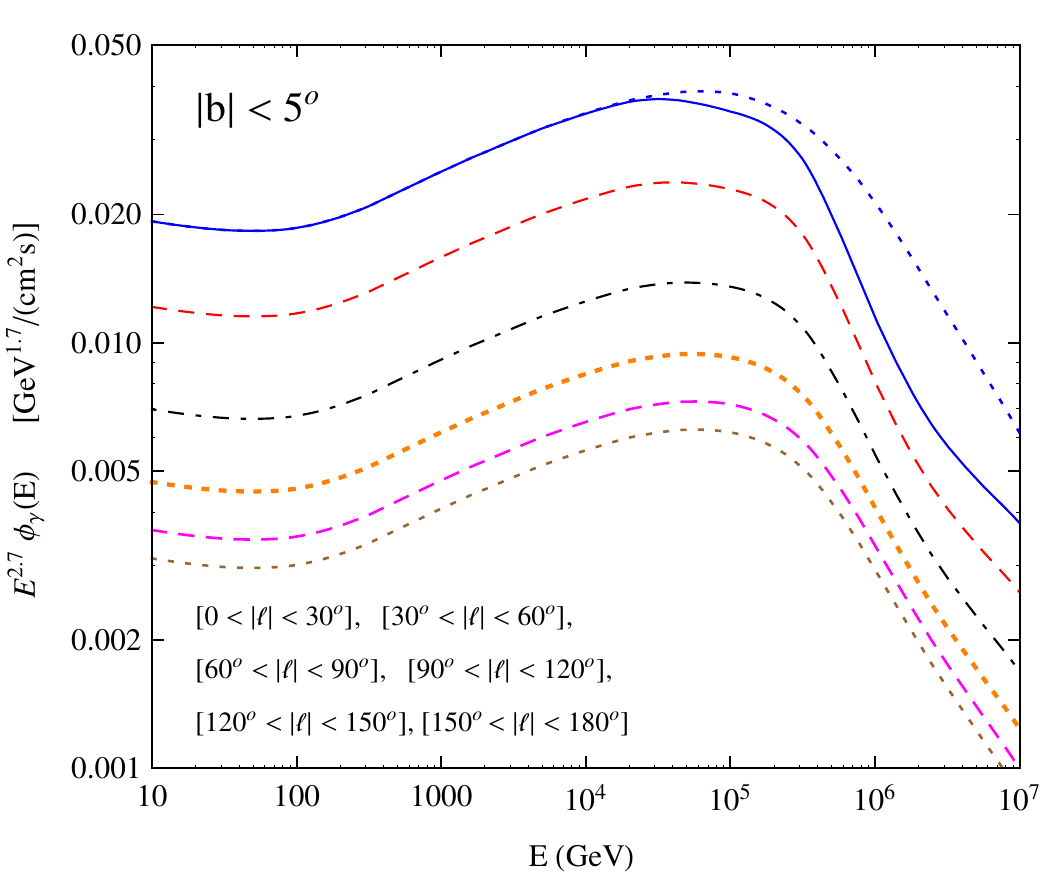}

\vspace{0.25 cm}
\includegraphics[width=8.0cm]{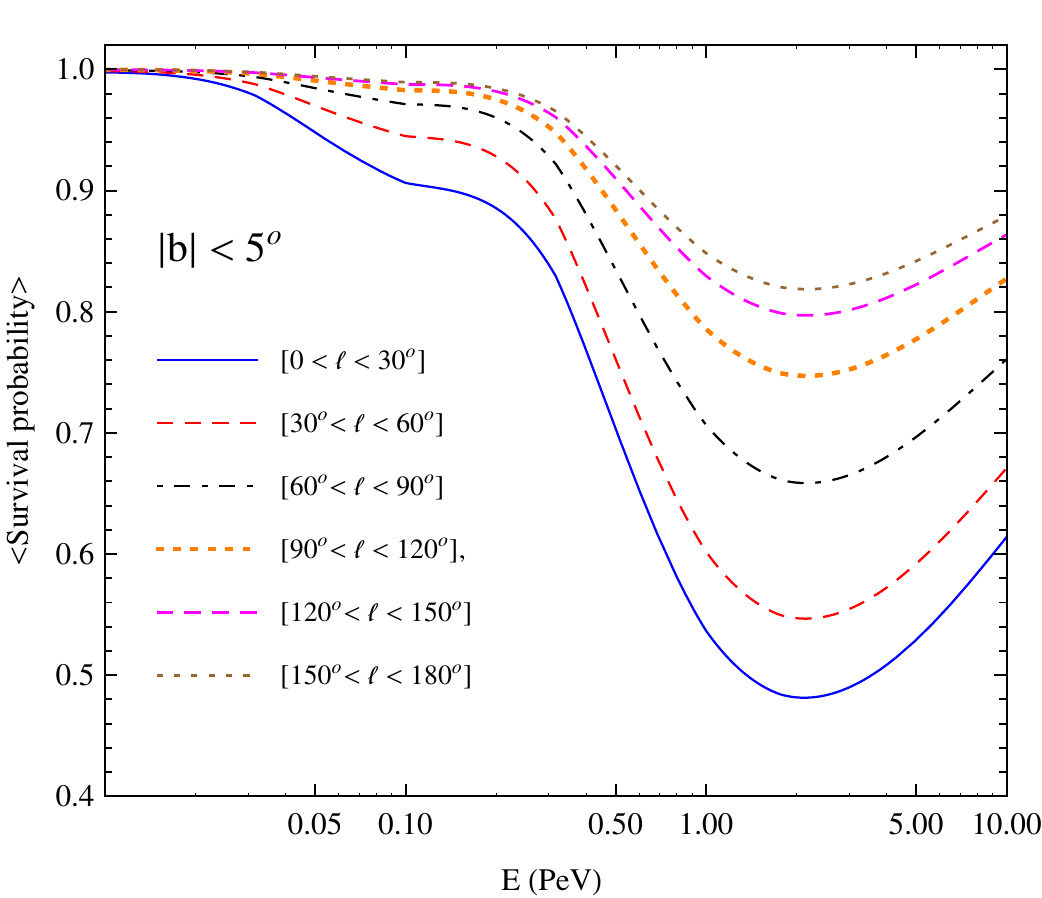}
\end{center}
\caption {\footnotesize
Top panel: 
gamma ray flux as a function of energy, according to the model where the emission is factorized.
The flux is integrated in latitude in the range $|b| < 5^\circ$ and in six longitude intervals.
From top to bottom:
$|\ell| < 30^\circ$,
$30^\circ < |\ell| < 60^\circ$,
$60^\circ < |\ell| < 90^\circ$,
$90^\circ < |\ell| < 120^\circ$,
$120^\circ < |\ell| < 150^\circ$ and
$150^\circ < |\ell| < 180^\circ$.
Absorption effects are included. For the region
$|\ell| < 30^\circ$ the flux calculated neglecting
absorption is also reported (as a dotted line).
Bottom panel: average survival probability of gamma rays
in the same six angular regions.
\label{fig:probe0}}
\end{figure}

To test the factorization hypothesis encoded in Eq.~(\ref{eq:phi-factorization})
one needs to study the diffuse gamma ray in a broad energy range.
The results of a calculation of the gamma ray up to very high energy 
performed assuming the validity of the factorization hypothesis are shown in Fig.~\ref{fig:probe0}.
The top panel in the figure shows the gamma ray flux integrated in different
angular regions of the Galactic plane ($|b| < 5^\circ$) and plotted versus the energy.
The absolute value of the flux depends on the
angular region, being largest for regions toward the Galactic Center, and decreasing
for larger longitudes $|\ell|$, however the spectral shape of the flux
remains approximately equal for $E\lesssim 30$~TeV. At higher energy the
absorption effects distort the spectra in an angle dependent way.
The angle dependence of the absorption effects can be 
seen more clearly in the bottom panel of Fig.~\ref{fig:probe0} that shows
the average survival probability of the gamma rays, plotted as
a function of $E_\gamma$, for the different angular regions.

The spectral distortion exibits a structure that reflects the fact
that the absorption is generated by the interactions on two main components
of the radiation field, dust emitted infrared radiation that is most effective for
$E \simeq 100$~TeV, and the cosmic microwave background radiation (CMBR),
that is most effective for $E \simeq 2$~PeV (see discussion in \cite{Vernetto:2016alq}).
The distortion pattern that is formed on the spectrum is qualitatively similar
in different angular regions, however the amount of absorption is largest for
directions toward the Galactic Center and minimum for directions toward the Anticenter.
This can be easily understood, noting that the flux in directions toward the
center has its origin in points that are on average further away from the Earth.

\begin{figure}[hbt]
\begin{center}
\includegraphics[width=8.0cm]{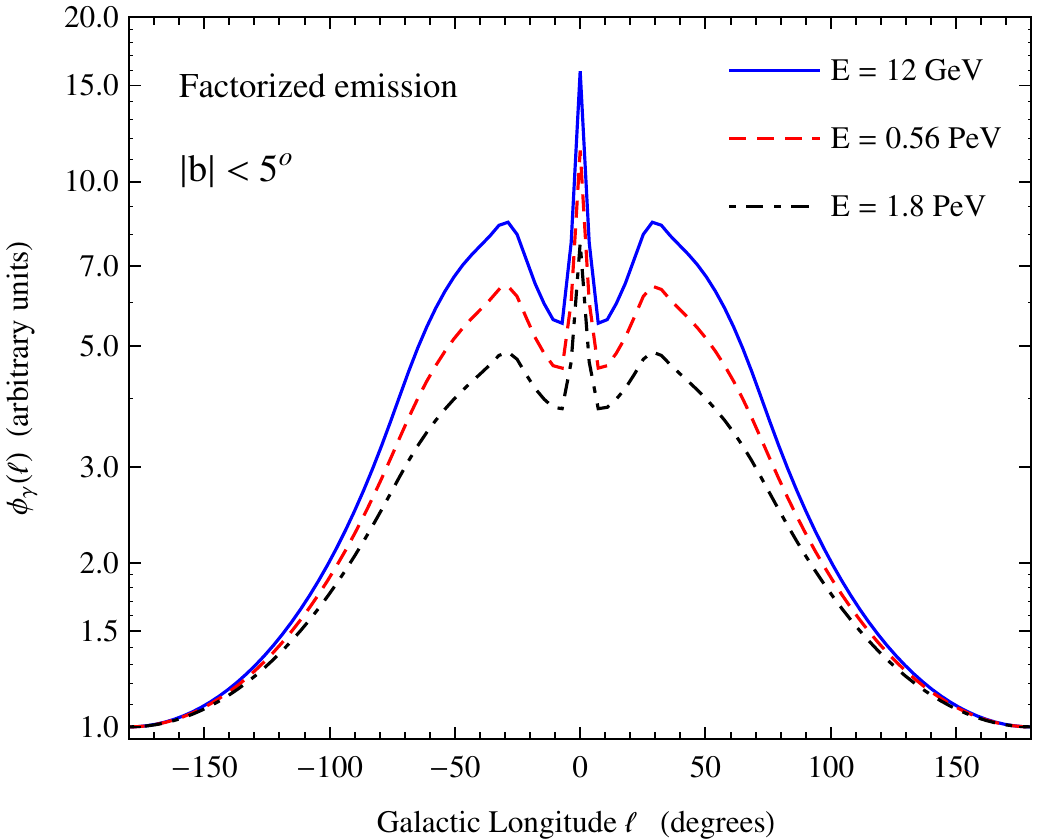}
\end{center}
\caption {\footnotesize
Longitude distribution of the gamma ray flux calculated in a model
where the emission is factorized.
The flux is shown for three values of the energy 
and it is rescaled to have value unity for $\ell = \pm 180^\circ$
to make easier a comparison of the shapes of the distributions.
The difference in shape for the three distributions is entirely due to 
energy dependent absorption effects.
\label{fig:ang-abs1}}
\end{figure}

The absorption effects are illustrated in a complementary way
in Fig.~\ref{fig:ang-abs1} that shows the longitude dependence of the flux,
after integration in latitude in the range
$|b| < 5^\circ$, for three values of the energy:
$E \simeq 12$~GeV, where absorption is
completely negligible, $E \simeq 0.56$~PeV where absorption is significant,
and $E \simeq 1.8$~PeV where absorption is largest.
In the figure the gamma ray flux is rescaled to have unit value at $|\ell| = 180^\circ$, for a better
visualization of the absorption effects. As it is intuitively obvious,
the flux in directions toward the Galactic Center is more suppressed
by absorption than the flux toward the Anticenter.

It should be noted that the effects of absorption remain always smaller than
a factor of order two, even in the case where they are most important,
that is for $E$ of order 1--3~PeV, and directions toward the Galactic Center. 

\section{Model 2: space dependent CR spectra} 
\label{sec:non-factorized}

If one or more of the conditions listed in section~\ref{sec:factorized} are non satisfied,
the spectra of cosmic rays can have a space dependent shape.
Most models for the $e^\mp$ spectra assume that this is the case because
the particles can lose a significant amount of energy propagating from the sources
to distant regions of the Galaxy.
For protons and nuclei,
that have a much smaller $|dE/dt|$, energy loss effects are
are expected to be negligible, but a space dependence of the spectral shape can
be generated by other mechanisms.

\begin{figure}[hbt]
\begin{center}
\includegraphics[width=8.0cm]{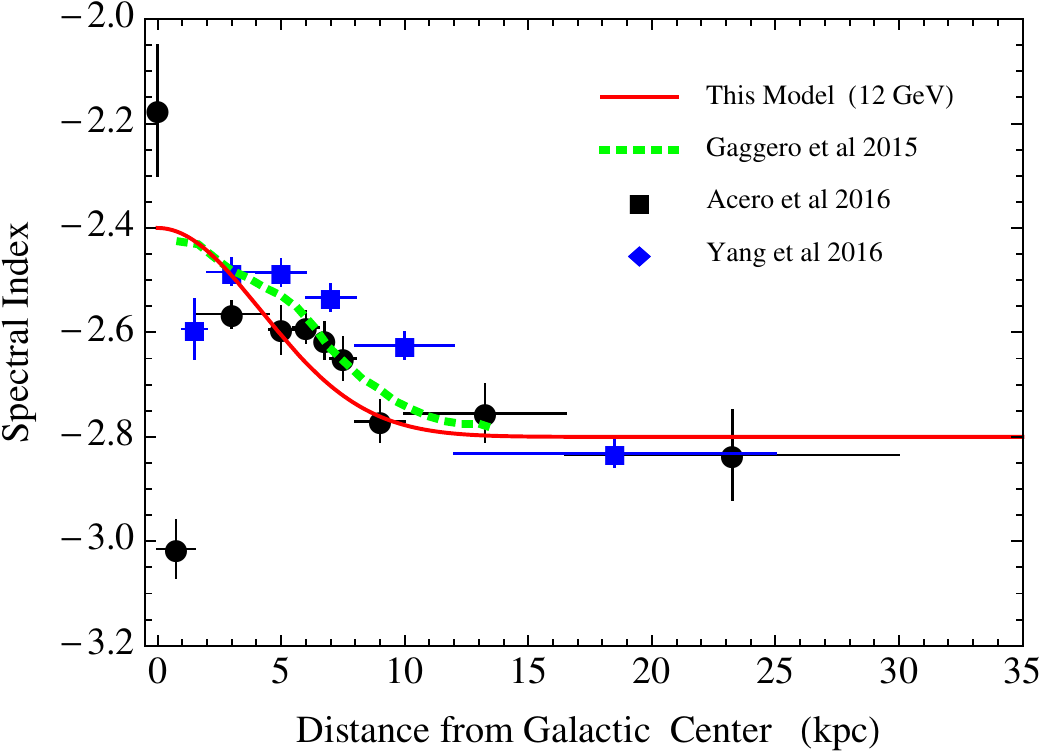}
\end{center}
\caption {\footnotesize
Spectral index of the gamma ray emission as a function
of the distance from the Galactic Center
for points on the Galactic plane.
The points are the estimates by
Acero et al. \cite{Acero:2016qlg} 
and Yang et al. \cite{Yang:2016jda}.
The dashed line is from Gaggero et al. \cite{Gaggero:2014xla}.
The solid line is the model discussed in this paper
for $E = 12$~GeV.
\label{fig:index1}}
\end{figure}

Some recent analysis of the Galactic diffuse flux \cite{Gaggero:2014xla,Acero:2016qlg,Yang:2016jda}
conclude that there is some evidence for the fact that
cosmic rays in the central part of the Galaxy
have a harder spectrum than what is observed at the Earth, while
cosmic rays in the periphery of the Galaxy are (moderately) softer.
This effect can be described as a space dependence of the 
spectral index of the gamma ray emission.
Fig.~\ref{fig:index1} shows some estimates of the
dependence of the gamma ray emission spectral index
on the distance from the Galactic Center.
It has to be noted that a crucial problem in establishing the existence of these effects
is to take into account the contribution of unresolved discrete
Galactic sources. This problem will be discussed 
in section~\ref{sec:point-sources}.

Aiming at the construction of a model as simple as possible
we have assumed that the spectral index at the
refence energy $E^* = 12$~GeV has a space dependence specified by the functional form: 
\begin{equation}
 \alpha_\gamma^{*} (r,z) = \alpha_{\rm max} - (\alpha_{\rm max} -\alpha_{\rm min})
 \; e^ {-r^2/(2 \sigma_r^2)} 
 \; e^ {-z^2/(2 \sigma_z^2)} 
\end{equation}
The parameters in this equation have been chosen so that
the spectral index at the Galactic Center has the value $\alpha_{\rm min} = 2.4$,
while the index at large $r$ is $\alpha_{\rm max} = 2.8$.
The $r$ dependence is described by the parameter $\sigma_r = 4.17$~kpc.
This corresponds to a spectral index in the vicinity of the solar system
$\alpha^*_\odot \equiv \alpha^* (r_\odot,0) = 2.75$.
The studies of \cite{Gaggero:2014xla,Acero:2016qlg,Yang:2016jda}
have not observed a $z$ dependence of the spectral index. This
can be expressed as a lower limit $\sigma_z \gtrsim 0.4$~kpc.

The gamma ray emission in the model is then described by the expression:
\begin{equation}
 q_\gamma (E, \vec{r}) = q_\gamma^{\rm loc}(E) \; f_0 (\vec{r}) \;
 \left [\frac{n(\vec{r})}{n(\vec{r}_\odot)} \right ] \; \left ( \frac{E}{E^*} 
 \right )^{-[\alpha^*(\vec{r}) - \alpha_\odot^*]}
\label{eq:q-non-factorized} 
\end{equation}
It is straightforward to see that this model is identical to the one
discussed in section~\ref{sec:factorized} for $E = E^*$ when the last factor
in equal to unity, however for $E > E^*$ this model and the factorized model
of Eq.~(\ref{eq:q-factorized}) start to diverge.

\begin{figure}[hbt]
\begin{center}
\includegraphics[width=8.0cm]{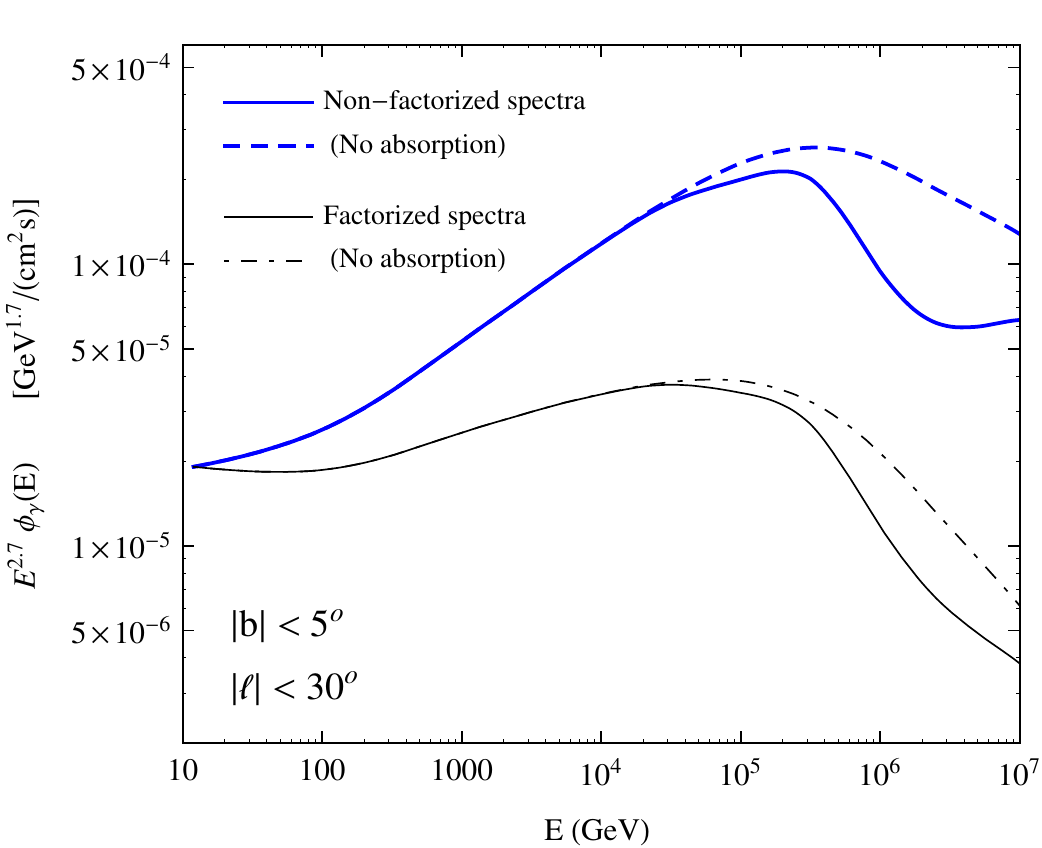}

\vspace{0.25 cm}
\includegraphics[width=8.0cm]{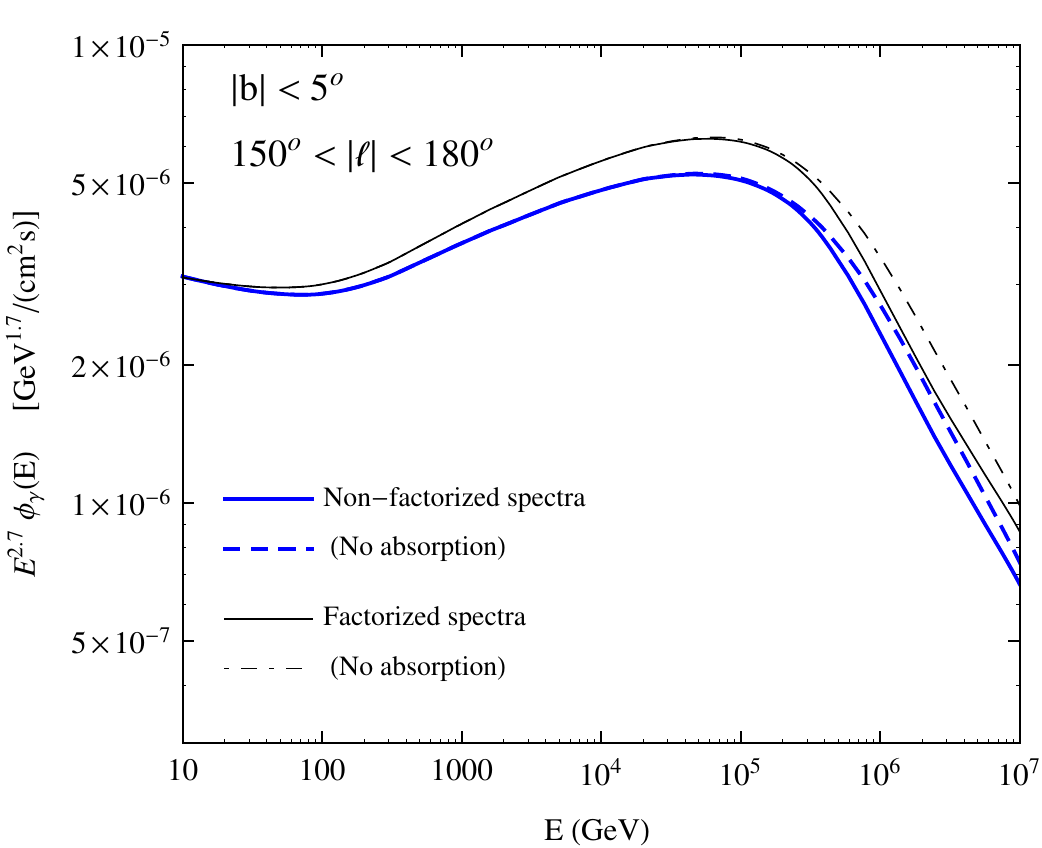}
\end{center}
\caption {\footnotesize
Energy spectra of diffuse gamma rays according to different
models of emission.
Thin lines: model
where the emission is factorized.
Thick lines: model where the factorization is not valid.
The solid (dashed) lines show the flux calculated including
(neglecting) the effects of gamma ray absorption.
Top panel: the flux is integrated in the angular region 
$|b| < 5^\circ$, $|\ell| < 30^\circ$.
Bottom panel: the flux is in the angular region $|b| < 5^\circ$, $150 < |\ell| < 180^\circ$.
\label{fig:flux-center}}
\end{figure}

This effect is illustrated in Fig.~\ref{fig:flux-center}
that shows the gamma ray energy spectra calculated in the two models
after integration in two angular regions toward the Galactic Center (top panel)
and the Anticenter (bottom panel).
For the angular region around the Galactic Center
($|b| < 5^\circ$ and $|\ell| < 30^\circ$), the spectrum calculated in the
non--factorized model is significantly harder, and the ratio
between the two models grows with energy.
The non--factorized model becomes a factor of ten larger for $E \approx 1$~PeV. On the contrary, for the
angular region around the Galactic Anticenter, the non--factorized model has a spectrum
that is slightly softer.
In this case the difference between the models is smaller
(of order 20\% for energies of order 1~PeV).

\begin{figure}[hbt]
\begin{center}
\includegraphics[width=8.0cm]{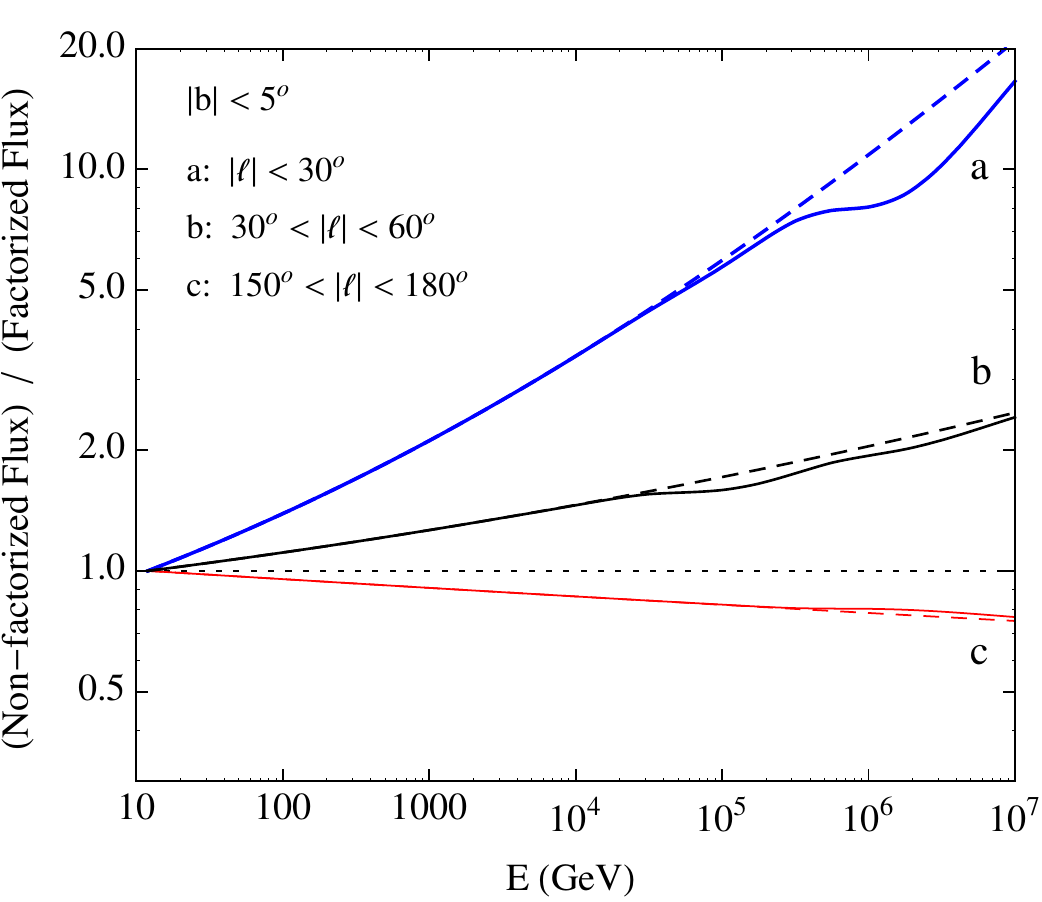}
\end{center}
\caption {\footnotesize
 Ratio between the gamma ray fluxes calculated according to
 the non--factorized and the factorized emission models.
The ratio is shown as a function of the gamma ray energy,
after integration in different longitude regions, for the latitude $|b| < 5^\circ$. 
The solid (dashed) lines include (neglect) the effects of absorption.
\label{fig:flux-energy-ratio}}
\end{figure}

These points are also illustrated in Fig.~\ref{fig:flux-energy-ratio}, that shows
the ratio of fluxes calculated in the two models for the two regions discussed above,
and also a third intermediate region ($|b| < 5^\circ$ and $30^\circ \le |\ell| < 60^\circ$).
In this third region the non--factorized model is moderately harder,
with a ratio of order two in the PeV energy range.

\begin{figure}[hbt]
\begin{center}
\includegraphics[width=8.0cm]{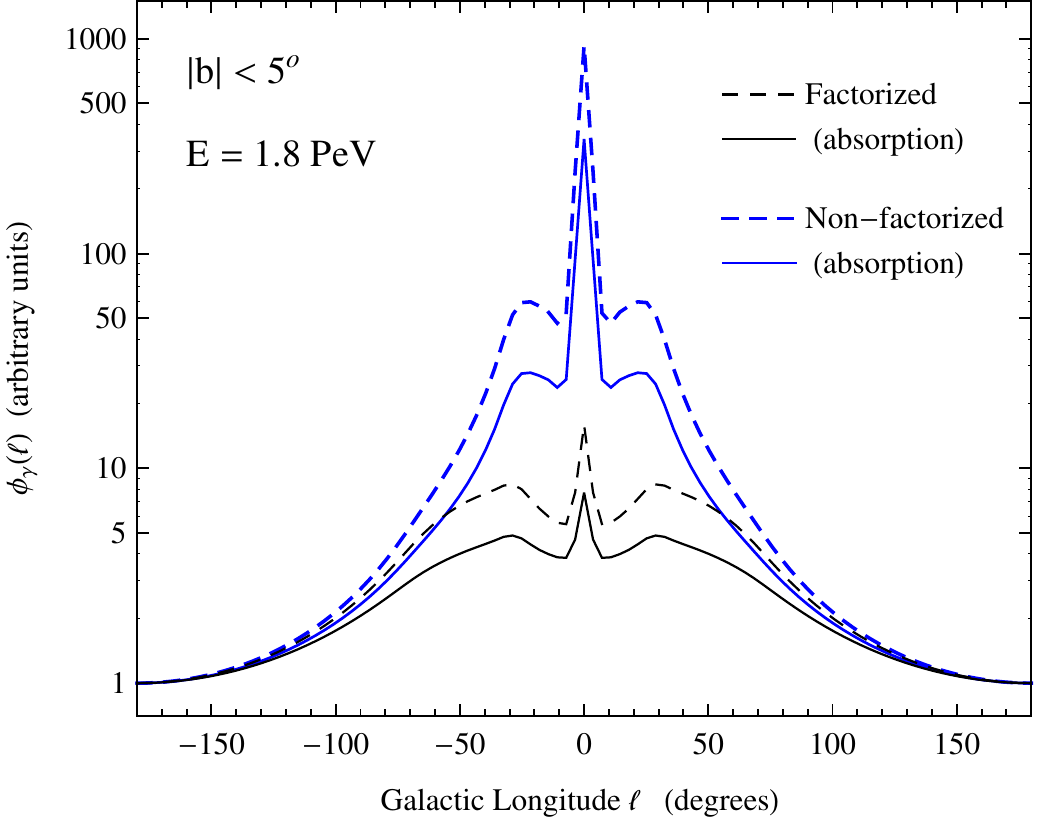}

\vspace{0.9 cm}
\includegraphics[width=8.0cm]{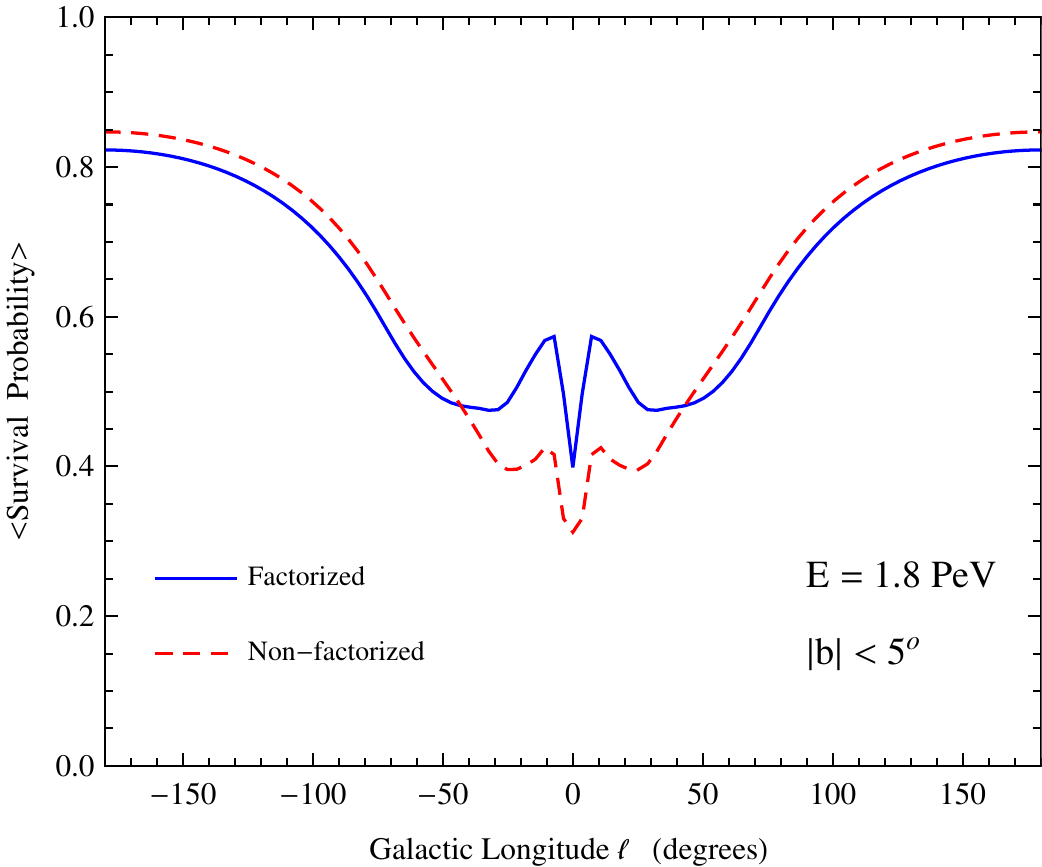}
\end{center}
\caption {\footnotesize
Top panel: longitude distribution of the gamma ray flux at $E = 1.8$~PeV, integrated in the latitude range $|b| < 5^\circ$. 
The flux is shown for both our models (factorized and non--factorized emissions),
including and neglecting the effects of absorption.
Bottom panel: average survival probability for gamma rays of energy $E = 1.8$~PeV, averaged in the latitude interval $|b| < 5^\circ$, as a function of the Galactic longitude, according to our two models. 
\label{fig:ang-abs2}}
\end{figure}

The same information can of course be obtained studying 
the shape of the angular distribution of the diffuse flux at different energies in the two models.
As discussed in the previous section, in a factorized model the angular distribution is energy
independent, except for absorption effects.
For a non--factorized model, such as the one we have constructed here,
the enhancement of the flux from directions toward the Galactic Center
becomes more and more significant with increasing energy.
This is illustrated in Fig.~\ref{fig:ang-abs2}, where the top panel shows the shapes of the
longitude distribution of the gamma ray flux at energy of 1.8~PeV, in the two models.
The ratio between the fluxes in the
directions around the Galactic Center and
Anticenter is one order of magnitude larger in the non--factorized model.

\begin{figure}[hbt]
\begin{center}
\includegraphics[width=8.0cm]{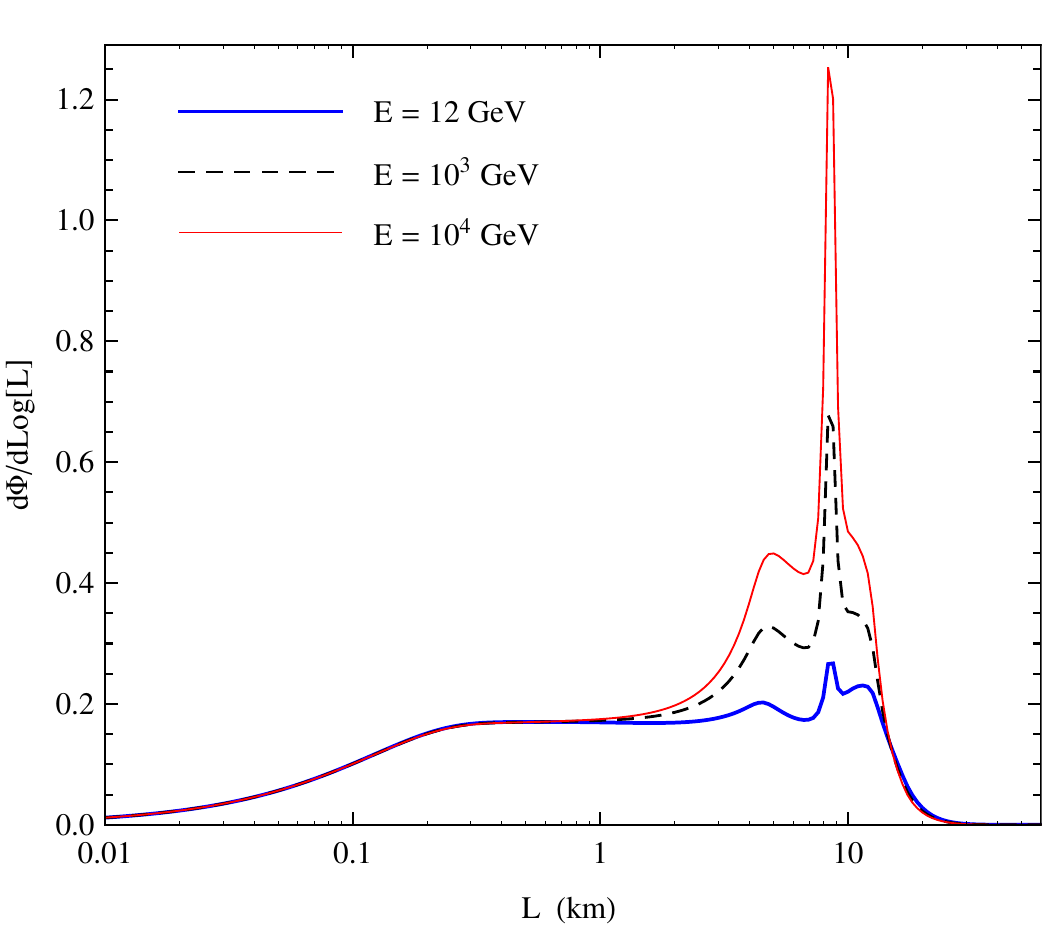}
\end{center}
\caption {\footnotesize
Distribution of the distance traveled
by photons of the diffuse Galactic emission (neglecting absorption effects).
The thick solid curve is the distribution 
at the reference energy $E^* = 12$~GeV.
The other curves are calculated for the energies
$E = 10^3$ and $10^4$~GeV,
according to the non--factorized model for the emission.
\label{fig:dist_diffuse}}
\end{figure}

The survival probabilities for the two models are shown in the bottom panel of
Fig.~\ref{fig:ang-abs2}.
The two probabilities are close to each other, but not identical
reflecting the difference in the space distribution of the emission.
This difference can be visualized inspecting Fig.~\ref{fig:dist_diffuse}
that shows the distribution of pathlength of the photons that form the
diffuse Galactic emission at the Earth.
The figure clearly shows how a very broad range
of pathlengths contribute to the diffuse flux. In the non--factorized model,
the contribution to the flux of points in the central region
of the Galaxy becomes enhanced with increasing energy.

\section{The IceCube neutrino signal}
\label{sec:icecube}
As discussed in the introduction, the IceCube neutrino telescope
has recently obtained evidence for the existence of a signal of high energy
events of astrophysical origin above the expected foreground of atmospheric $\nu$'s
\cite{Aartsen:2013jdh,Aartsen:2014gkd,Aartsen:2015rwa,Aartsen:2017mau}.
The signal is consistent with an isotropic
flux of extragalactic neutrinos, generated by the ensemble of
all (unresolved) sources in the universe.
The flavor composition of the events in the signal
(with the three flavors having approximately the same flux)
is consistent with the expected composition of
a flux generated by the standard mechanism of pion decay,
after taking into account flavor oscillations
(and averaging over a broad range of $\nu$ pathlengths).

\begin{figure}[hbt]
\begin{center}
\includegraphics[width=8.0cm]{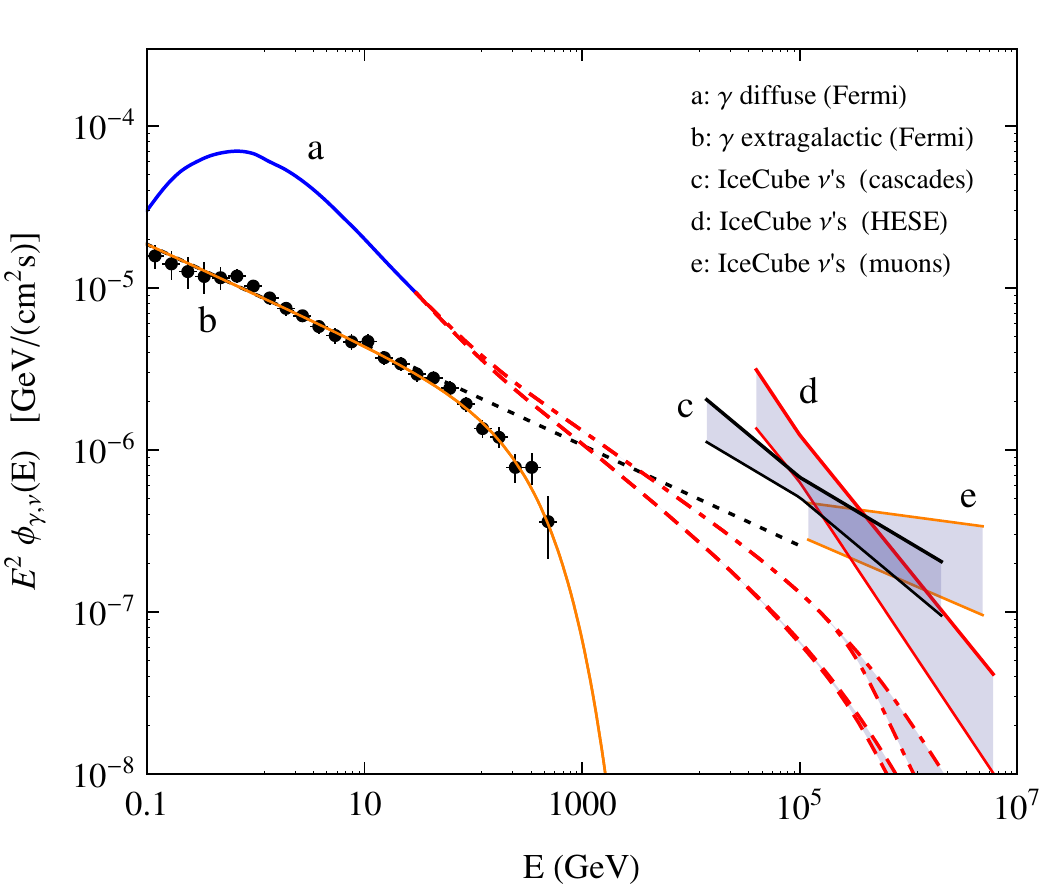}
\end{center}
\caption {\footnotesize Curve $a$ shows the diffuse Galactic gamma ray flux observed by Fermi
(solid line) and the extrapolations calculated under the assumptions of a factorized
(dashed line) and non--factorized (dot--dashed line) emission.
The Fermi measurement of the extragalactic gamma ray flux \protect\cite{Ackermann:2014usa}
is shown by as line $b$ (and with the points).
The shaded regions $c$ $d$ and $e$ show estimates of the neutrino flux obtained by IceCube
under the assumption of an isotropic extragalactic flux \protect\cite{Aartsen:2017mau}.
All fluxes are integrated over the entire sky.
 \label{fig:neutrino_gamma}}
\end{figure}

Power law fits to the neutrino energy spectrum
in the range $E_\nu \approx 30$--$10^4$~TeV,
performed under the hypothesis that
the signal is an isotropic extragalactic flux, 
have been recently presented by IceCube \cite{Aartsen:2017mau}
for different classes of events and are shown in Fig.~\ref{fig:neutrino_gamma}.

If the neutrinos of the IceCube signal are generated by a standard
production mechanism, the $\nu$ emission should be accompanied by
an emission of gamma rays with approximately equal spectral shape and normalization.
If the neutrinos are extragalactic, 
one does not expect to observe an associated high energy photon flux
because the gamma rays are (to a very good approximation)
completely absorbed during propagation.
On the other hand, if a significant fraction of the
$\nu$ signal is of Galactic origin, the corresponding gamma rays flux
is only partially absorbed and remains observable.

In Fig.~\ref{fig:neutrino_gamma} the IceCube fits to the neutrino
spectrum are shown together with the measurements of the
extragalactic and diffuse Galactic gamma ray
fluxes obtained by Fermi, and also with the extrapolations
of the diffuse Galactic flux (for the factorized and non--factorized models)
that are discussed in this paper.
Note that the figure shows angle integrated fluxes, and that the Galactic
gamma ray fluxes have a strong angular dependence.

The comparison of the $\gamma$ and $\nu$ fluxes
indicates that the IceCube signal is
significantly higher than the 
diffuse Galactic flux predicted on the basis
of ``natural'' extrapolations of the observations at lower energy,
even if one allows for the possibility
that the emission of gamma rays and neutrinos is harder
in the central part of the Galaxy.
Similar results for the diffuse Galactic
neutrino flux have been obtained by \cite{Pagliaroli:2016lgg}.

Stringent limits on the flux of astrophysical neutrinos from
the Galactic disk have been obtained by ANTARES \cite{Albert:2017oba}.

Several authors have however suggested that a significant fraction of
(or even the entire) IceCube signal is of Galactic origin.
This requires the introduction of some new mechanism
for $\nu$ production to explain the higher normalization and
the approximately isotropic angular distribution of the neutrino
signal.

For example, Esmaili and Serpico
have suggested that the neutrino
signal is generated by the decay of a dark matter particle
with a mass of a few PeV
\cite{Esmaili:2013gha,Esmaili:2014rma}.
In this case the space distribution of the
emission is proportional to the dark matter density.
Esmaili and Serpico adopt a Navarro--Frenk--White density profile:
\begin{equation}
\rho(r) \simeq \frac{\rho_0}{r/r_c \, (1 + r/r_c)^2} 
\end{equation}
with $r_c \simeq 20$~kpc and $\rho_0 \simeq 0.33$~GeV/cm$^{-3}$.

Taylor, Gabici and Aharonian \cite{Taylor:2014hya}
have suggested that the IceCube signal is generated by
the interactions of cosmic rays filling a very extended spherical halo around
the disk of the Milky Way with a size of order 100~kpc.
In the following we will model
the space dependence of the emission as a simple gaussian:
$q(r) \propto e^{-r^2/R_0^2}$ with $R_0 \simeq 58$~kpc
(so that $\sqrt{\langle r^2 \rangle } = \sqrt{3} \, R_0 \simeq 100$~kpc).

Lunardini et al. \cite{Lunardini:2013gva}
 have suggested that a significant fraction
of the IceCube signal is generated in the Fermi bubbles
\cite{Su:2010qj,Fermi-LAT:2014sfa}.
In the following we will model the bubbles as spheres
of radius $R_b \simeq 3.85$~kpc, with centers above and below
the Galactic Center at $z = \pm 3.89$~kpc.
The emission density in the Fermi bubbles grows with radius,
and has been fitted with the form $\propto 1/\sqrt{1- r^2/R_b^2}$.

Assuming that the gamma ray emission in the three models
listed above (dark matter decay, large halo and Fermi bubbles)
has the same spectrum for all points, it is straightforward to
compute the angular distribution of the emission and the absorption
probability as a function of the energy and direction.

\begin{figure}[hbt]
\begin{center}
\includegraphics[width=8.0cm]{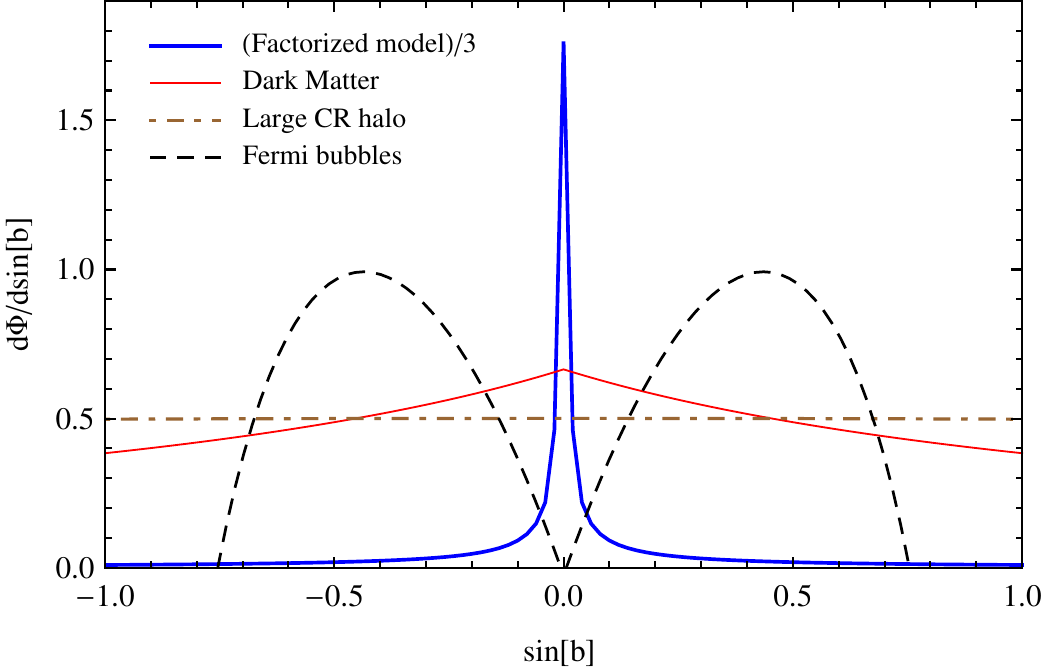}

\vspace{0.9 cm}
\includegraphics[width=8.0cm]{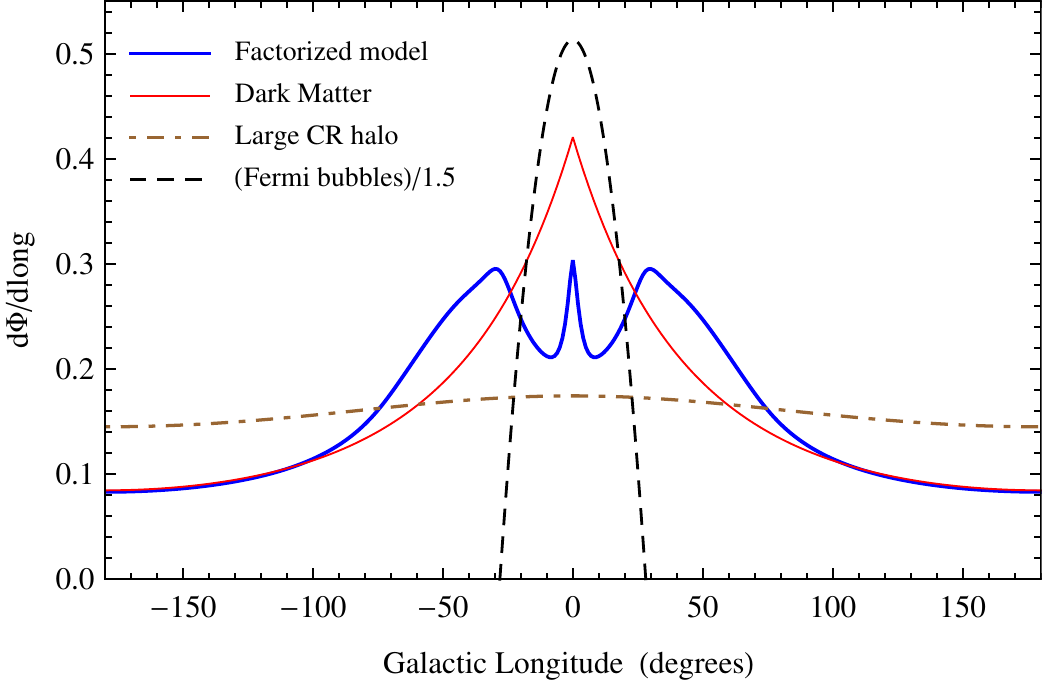}
\end{center}
\caption {\footnotesize
 Angular distributions of the Galactic gamma ray flux
 for different emission models:
 Galactic model with factorized emission, dark matter decay model
 \cite{Esmaili:2013gha,Esmaili:2014rma},
 large CR halo model \cite{Taylor:2014hya}
 and emission from the Fermi bubbles \cite{Lunardini:2013gva}.
 Top panel: Latitude ($\sin b$) distributions. Bottom panel: longitude distributions.
 The curves are calculated neglecting the absortpion effects and are
normalized to unit area (except where noted).
 \label{fig:gamma-angular}}
\end{figure}

The latitude and longitude distributions of the flux observable at
the Earth (calculated neglecting the effects of absorptions)
are shown in Fig.~\ref{fig:gamma-angular}
for the three models, together with the distribution of the factorized model discussed
previously.

\begin{figure}[hbt]
\begin{center}
\includegraphics[width=8.0cm]{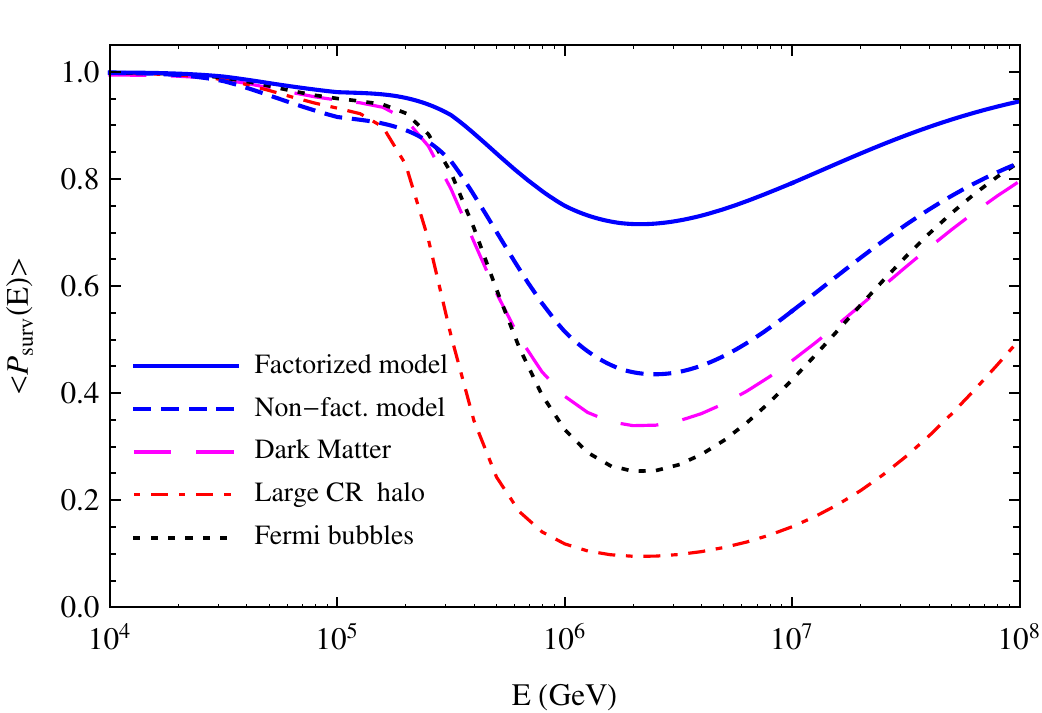}

\vspace{0.8 cm}
\includegraphics[width=8.0cm]{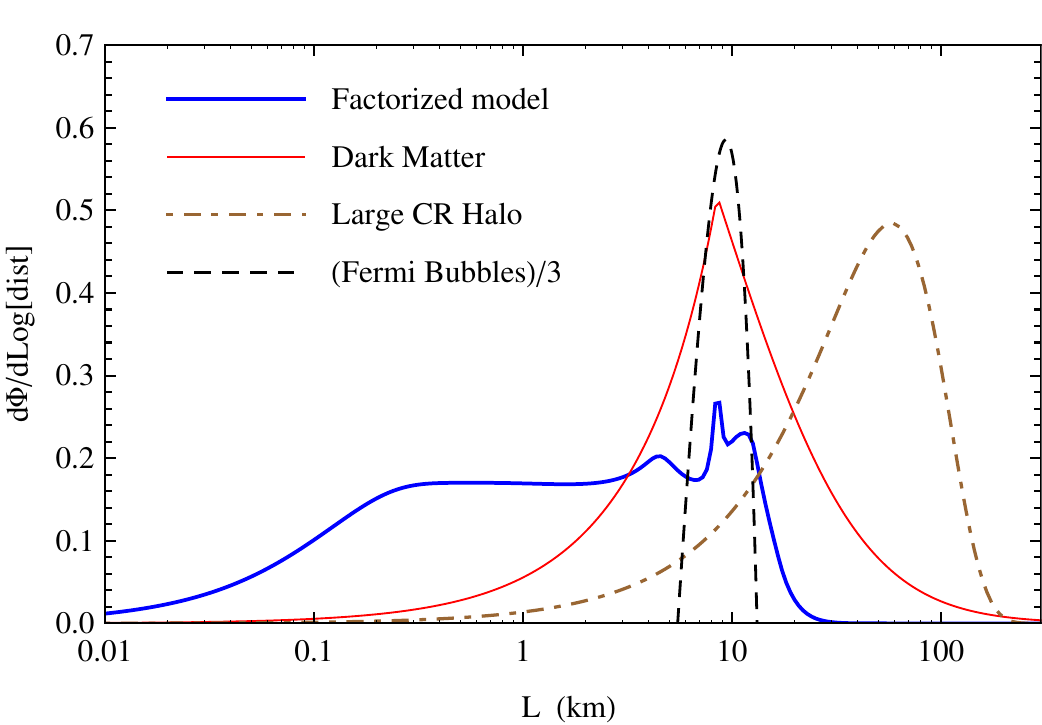}
\end{center}
\caption {\footnotesize
 Top panel: average survival probability of gamma rays
 (after integration over the entire sky) for different emission models.
 Bottom panel: distribution of the gamma ray pathlength for the same models. 
 The distributions are calculated after integration over the entire solid angle,
 neglecting the effects of absorption, 
 and are normalized to unity (or 1/3 for the Fermi bubbles case).
 \label{fig:proball}} 
\end{figure}

The gamma ray survival probabilities for the three non--standard
models (averaged over all directions) are shown in the top panel of Fig.~\ref{fig:proball}.
The probability of absorption is significantly
larger than for conventional emission models,
reflecting the fact that the average pathlength of the photons is longer.
The pathlength distributions for the different models are shown
in the bottom panel of Fig.~\ref{fig:gamma-angular}.

It should be noted that the knowledge of the pathlength distribution is not sufficient
to estimate exactly the gamma ray absorption effects
because the absorption probability per unit length has a non trivial space dependence.
However, the most important component of the target radiation fields
(the CMBR) is homogeneous and isotropic, and therefore 
it is possible to quickly obtain
a reasonably accurate estimate of the absorption probability
considering only the CMBR and the pathlength distribution.

\section{Galactic point sources}
\label{sec:point-sources}
The measurement of the diffuse Galactic flux requires the subtraction of
the extragalactic flux and of the contribution of all
Galactic sources.

The extragalactic flux is (to a very good approximation)
isotropic and can be accurately measured in polar regions
(at large latitudes $|b|$) where the Galactic foreground is small and then subtracted
also in angular regions where the Galactic flux is dominant.
The subtraction of the contribution of the Galactic sources is more problematic,
because it requires an estimate of the contribution of {\em unresolved} sources.

Many discrete (point--like and extended) gamma ray sources have been identified
by Fermi in the 0.1--1000~GeV energy range \cite{Acero:2015hja,TheFermi-LAT:2017pvy},
and at higher energy by Cherenkov telescopes and air shower arrays
\cite{tevcat,hesscat,Bartoli:2013qxm,Abeysekara:2017hyn}.

\begin{figure}[hbt]
\begin{center}
\includegraphics[width=8.0cm]{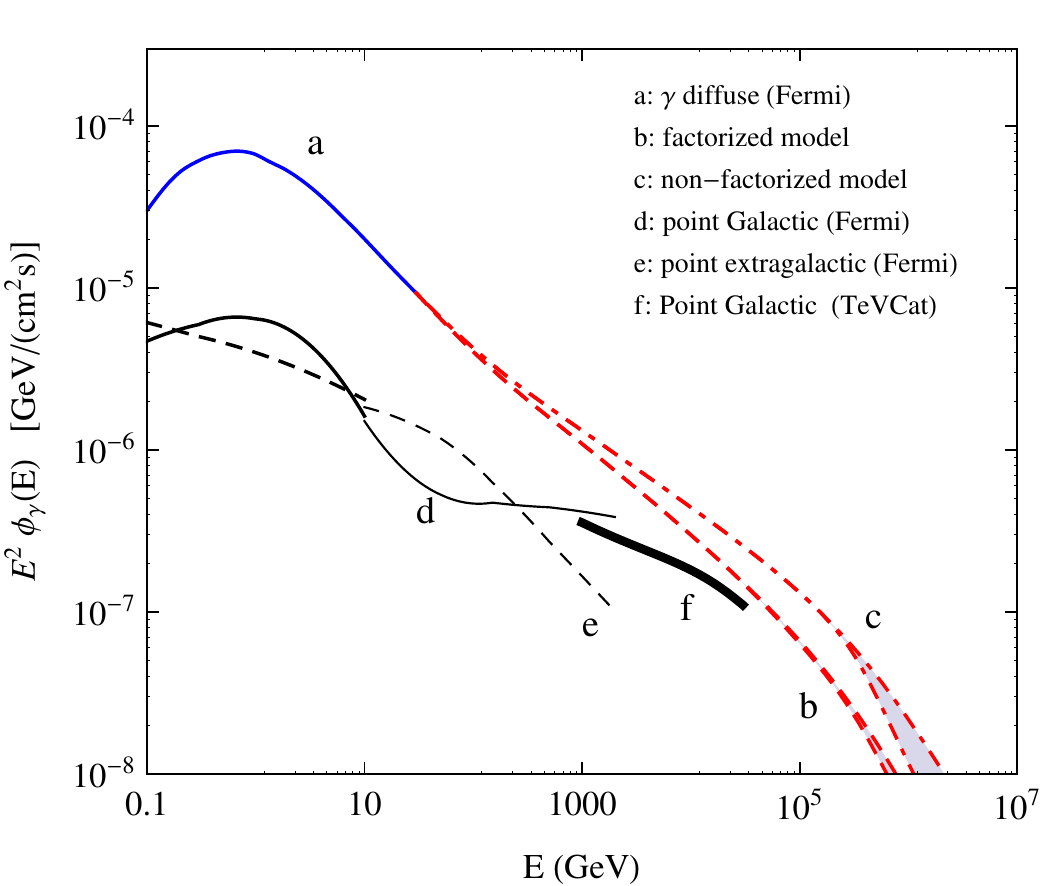}

\vspace{0.5 cm}
\includegraphics[width=8.0cm]{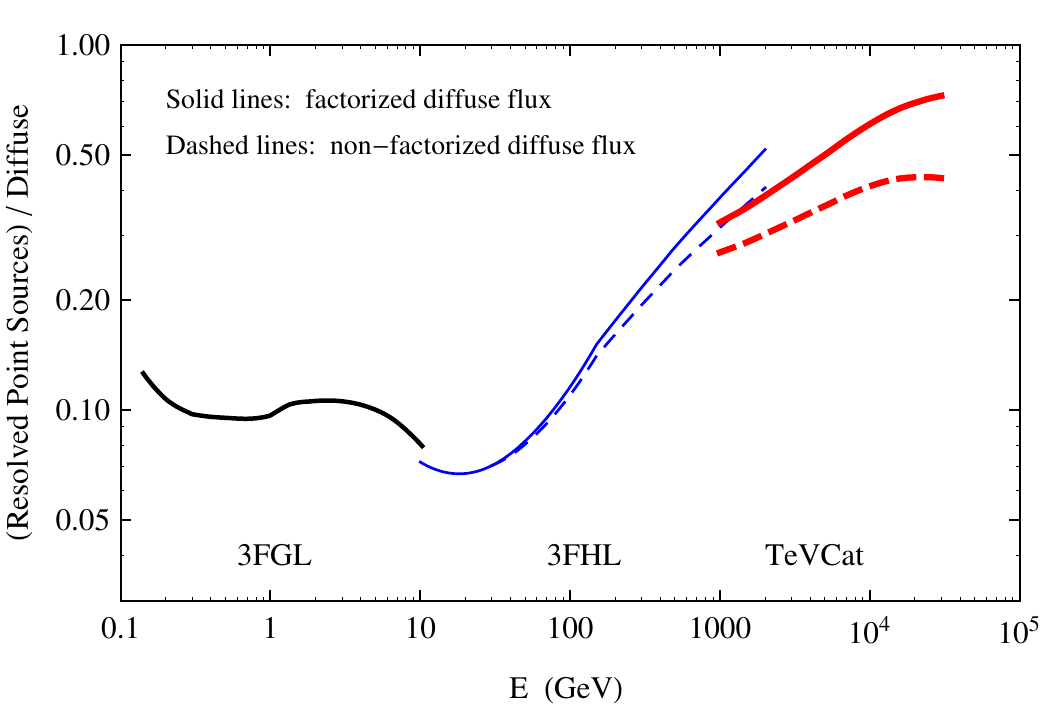}
\end{center}
\caption {\footnotesize 
Top panel, line $a$: Galactic diffuse flux
measured by Fermi; line $b$ and $c$: 
high energy extrapolations of the diffuse flux
according to our models (with the effects of absorption
visible for $E > 10^5$~GeV); 
line $d$: flux of the ensemble of Galactic sources
identified by Fermi in the 3FGL (thicker line) and 2FHL (thinner line) catalogs;
line $e$: flux of the ensemble of Fermi extragalactic sources;
line $f$: sum of the flux of all Galactic sources in the TeVCat catalog.
Bottom panel: ratio $\Phi_{\rm point}/\Phi_{\rm diffuse}$ between the sum of
the fluxes of all resolved Galactic sources and the
total Galactic diffuse flux. The ratio is calculated
for both the factorized and non--factorized models of the diffuse flux.
\label{fig:point_sources}}
\end{figure}

The cumulative flux of all detected Galactic sources
estimated from the published source catalogs is shown in the 
top panel of Fig.~\ref{fig:point_sources} and is compared to our extrapolations of
the diffuse Galactic flux.
In the case of the two Fermi catalogs
(3FGL \cite{Acero:2015hja} and 3FHL \cite{TheFermi-LAT:2017pvy})
the separation of Galactic and extragalactic sources is performed statistically.
The flux from extragalactic sources is estimated from a study of
the sources at large latitude ($|\sin b| > 0.025$). This contribution is then
assumed to be isotropic and subtracted from the flux of all sources
in the Galactic equatorial region to estimate the Galactic component. 
To estimate the flux at higher energy we have summed the flux of 116 sources with
$|b| < 10^\circ$ in the TeVCat online catalog \cite{tevcat}.

The bottom panel of Fig.~\ref{fig:point_sources} gives the ratio
$\Phi_{\rm point} / \Phi_{\rm diffuse}$ between the sum of all discrete resolved sources and the diffuse
Galactic flux.
For the diffuse flux we use the Fermi template for $E < 12$~GeV and
at higher energy the two extrapolations calculated in this work.

The important point of Fig.~\ref{fig:point_sources} is that
the flux of the ensemble of the detected discrete sources has a harder spectrum than
the diffuse flux, and the ratio
$\Phi_{\rm point}^{\rm resolved} (E) / \Phi_{\rm diffuse} (E)$
between the flux of the ensemble of all resolved sources
and the diffuse flux grows with energy 
from a value of order 0.06 at $E \simeq 10$~GeV to a value of order 0.3--0.5 at 1~TeV.

The estimate of the contribution of unresolved discrete sources to the Galactic flux
requires models for the luminosity function and space
distribution of the sources. In the present work we will not address this problem,
but we can note here that it is likely that also the
ratio $\Phi_{\rm point}^{\rm unresolved} (E) / \Phi_{\rm diffuse} (E)$ will grow
with energy.

Disentangling the contributions from gamma ray emission in interstellar space
(the diffuse component) from emission inside or in the vicinity
of sources is therefore an important problem, and an incorrect estimate of the
flux of unresolved sources can lead to incorrect conclusions.
One should also note that the contribution of unresolved sources
is expected to have a non trivial angular dependence, with most of the flux
arriving from directions toward the Galactic Center.

The problem of separating the ``source'' and ``diffuse'' components of the Galactic
emission is likely to become more difficult with increasing $E$.
At very high energy, when cosmic rays escape rapidly from the sources and then from the Galaxy,
the separation of the two components could in fact become impossible.

\section{Measurements of the diffuse gamma ray flux at high energy}
\label{sec:observations}

Some measurements of the TeV diffuse Galactic gamma ray flux have
been obtained in the recent past by high altitude air shower detectors located in
the Northern hemisphere.

After a long collection of upper limits, the first measurement
of the diffuse Galactic emission has been
reported by the Milagro detector \cite{Atkins:2005wu}
that measured the flux from the Galactic plane region
$\ell=40^{\circ}$--$100^{\circ}$ and $|b| < 5^{\circ}$
at energies above 3.5 TeV. 
The measurement was higher than the expectations, suggesting the possible
existence of a ``TeV excess'', perhaps connected to the ``GeV excess''
previously reported by EGRET (later recognized as an instrumental effect). 
It has been later suggested that this measurement
could include the contributions of several discrete sources
and therefore overestimate the diffuse flux \cite{Bartoli:2015era}.

\begin{figure}[hbt]
\begin{center}
\includegraphics[width=8.0cm]{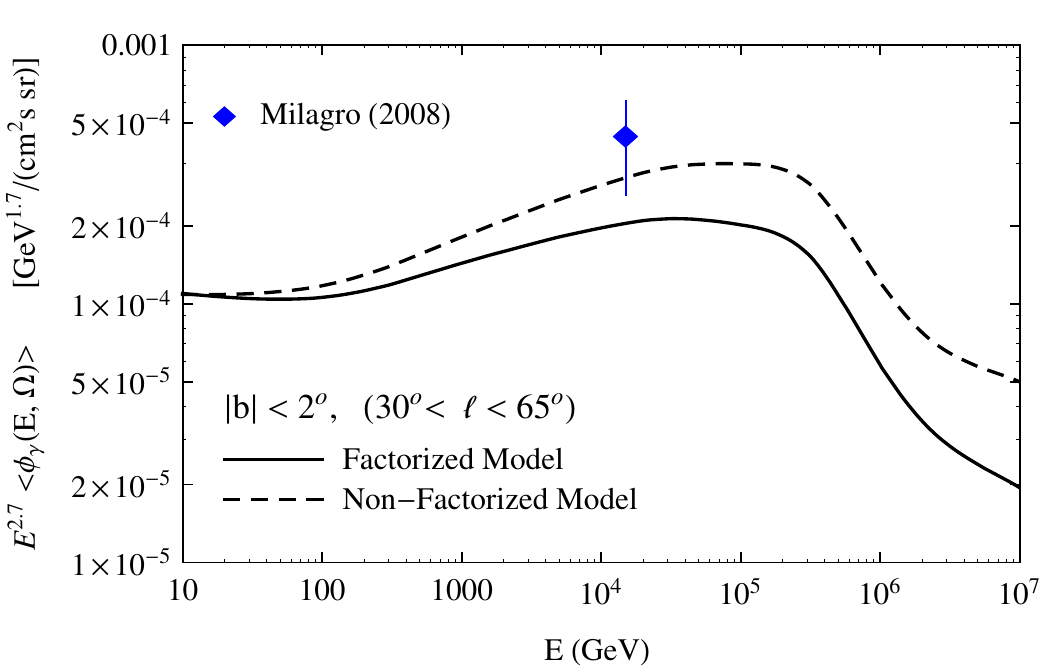}

\vspace{0.35 cm}
\includegraphics[width=8.0cm]{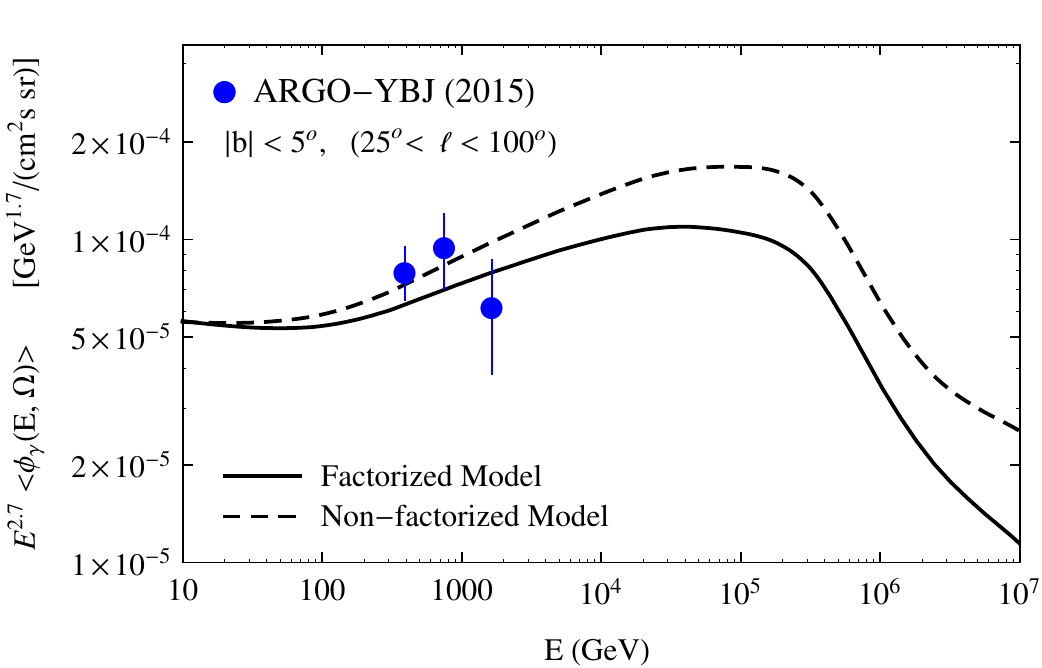}

\vspace{0.35 cm}
\includegraphics[width=8.0cm]{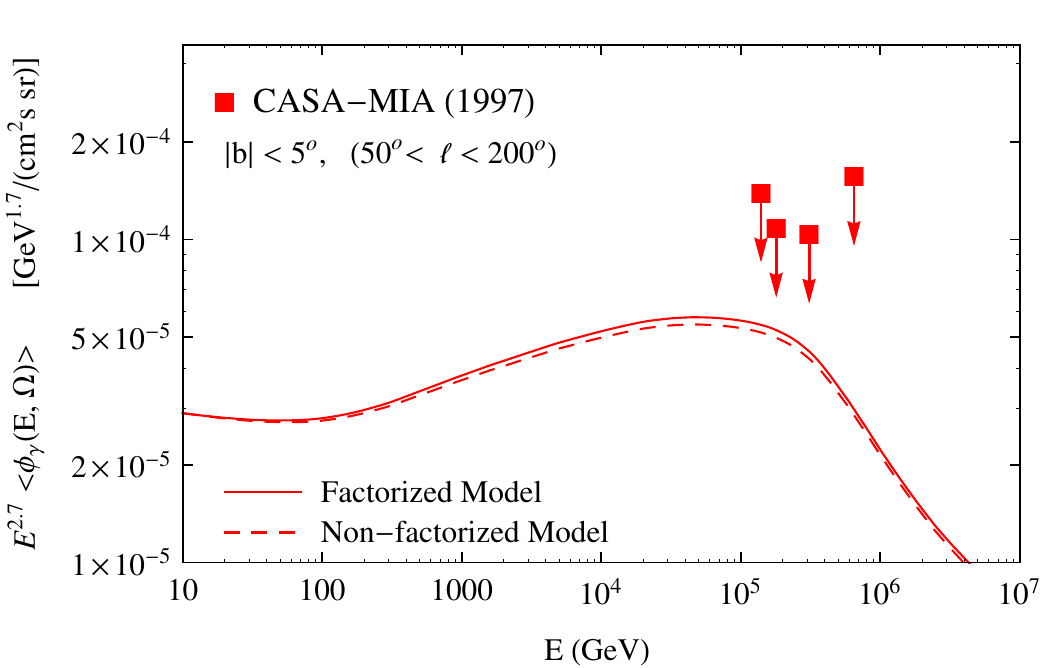}
\end{center}
\caption {\footnotesize
Measurements of the diffuse gamma ray flux at high energy, compared
to the average flux according to our two models,
calculated for the same angular region of the data. 
The three panels show the measurement of
Milagro \protect\cite{Abdo:2008if} (top),
ARGO--YBJ \protect\cite{Bartoli:2015era} (middle) and
CASA--MIA \protect\cite{Borione:1997fy} (bottom).
\label{fig:data-gamma}}
\end{figure}

The Milagro telescope has later measured the diffuse
flux at a median energy of 15~ TeV in the region
$\ell = 30^{\circ}$--$65^{\circ}$ and $|b| < 2^{\circ}$ \cite{Abdo:2008if}.
This measurement is shown in the top panel of 
Fig.~\ref{fig:data-gamma} where it is 
compared with the predictions of our models for the same angular region.
The Milagro observation is consistent
with both the factorized and non--factorized models,
but the second one seems to be slightly favoured.

The ARGO--YBJ detector has measured the energy spectrum of
the diffuse Galactic emission in the region $\ell=25^{\circ}$--100$^{\circ}$ and $|b|< 5^{\circ}$
for energies in the interval between $\sim 350$~GeV and $\sim 2$~TeV \cite{Bartoli:2015era}.
To evaluate the diffuse emission, the known gamma ray sources 
in the region have been masked, however a small residual contribution from non resolved sources
cannot be excluded.
The ARGO--YBJ measurement is shown in the middle panel of
Fig.~\ref{fig:data-gamma} and compared with our models.
The data points are consistent with the predictions,
but also in this case the large error bars 
do not allow to discriminate between the two models
that in this energy and angular region 
give predictions that are close to each other.

At higher energy ($E \gtrsim 100$~TeV)
the available measurements of the diffuse emission
have only provided upper limits.
The most stringent results have been obtained by CASA--MIA,
and constrain both the isotropic emission \cite{Chantell:1997gs}
and the emission from the Galactic plane \cite{Borione:1997fy}.
The bottom panel of Fig.~\ref{fig:data-gamma} shows the CASA--MIA flux upper limits
for the Galactic region $\ell =50^{\circ}$--200$^{\circ}$ and $|b|< 5^{\circ}$,
in the energy interval between 140 and 1300~TeV, compared to our predictions.
The CASA-MIA limits are a factor 2--5 higher than our models (depending on the energy).
It can be noted that measurements in this angular region
do not allow to discriminate between the factorized and non--factorized models
because the two predictions are very close to each other.

In summary, the existing measurements
of the diffuse Galactic gamma ray flux above 1~TeV are consistent with our extrapolations of
the Fermi observations, 
but are not capable to discriminate between the two models discussed in this work.
Future measurements with improved sensitivity, and a more
complete coverage of the Galactic plane have however the potential
to reach more firm conclusions on this problem.

\subsection{Detector sensitivity}
The study of the diffuse gamma ray flux at very high energy is probably best performed
with air shower detectors with the capability to discriminate between electromagnetic
cascades generated by photons and hadronic cascades generated by protons and nuclei.

\begin{figure}[hbt]
\begin{center}
\includegraphics[width=8.0cm]{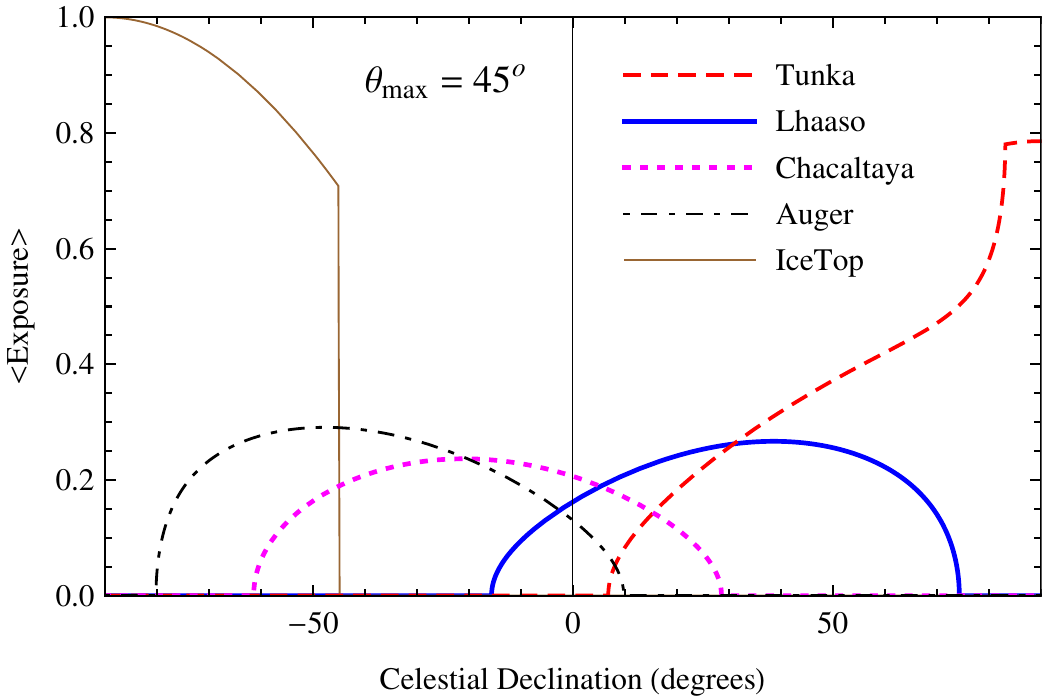}

\vspace{0.25 cm}
\includegraphics[width=8.0cm]{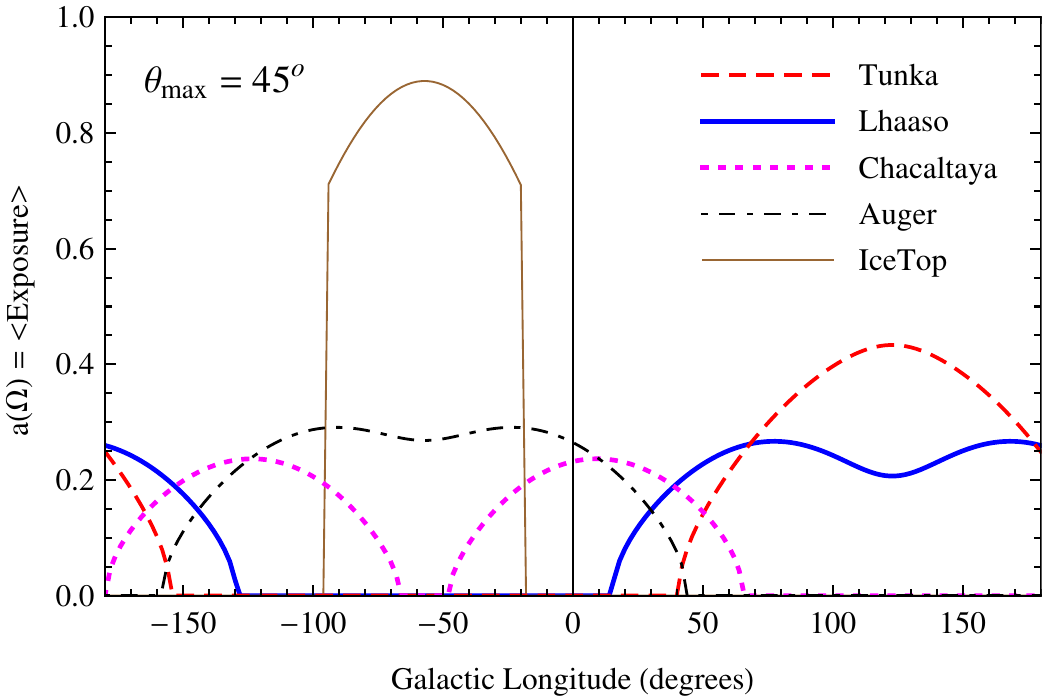}
\end{center}
\caption {\footnotesize
Top panel: average exposure for points in the sky with
celestial declination $\delta$.
Bottom panel: average exposure for points on the
Galactic equator, as a function of longitude.
The exposure is calculated for five detector locations:
Tunka ($\lambda = +51.8^\circ$),
LHAASO ($\lambda = +29.4^\circ$),
Chacaltaya ($\lambda = -16.4^\circ$),
Auger ($\lambda = -35.2^\circ$) and
IceTop ($\lambda = -90^\circ$). 
\label{fig:detection1}}
\end{figure}

For an order of magnitude estimate of the sensitivity of an air shower detector
to a diffuse gamma ray flux, one can note that the signal of gamma ray events
from a regions in the sky of angular size $\Delta \Omega$
around a point of celestial coordinates $\Omega = \{\delta, \alpha\}$ and
with energy in an interval of size $\Delta \ln E$ centered on $E$ is:
\begin{equation}
 S_{\gamma} \simeq E ~\phi_{\gamma} (E, \Omega) ~\Delta \ln E ~\Delta \Omega ~A \; T \; a(\delta, \lambda)
 ~\varepsilon_\gamma(E) ~.
\label{eq:nsignal}
\end{equation}
In this equation, $\phi_\gamma (E, \Omega)$ is the diffuse gamma ray flux,
$A$ is the detector area, $T$ is the observation time
(much longer than a sidereal day), $\varepsilon_\gamma(E)$ is the gamma ray detection
efficiency, and $a(\delta, \lambda)$ is an adimensional factor that takes into account
the visibility of the sky region under study at the detector geographical position.
The quantity $a(\delta, \lambda)$ takes into account the fraction of a sidereal day
that a point of celestial declination $\delta$ spends in the zenith angle range
($\theta_z < \theta_{\rm max}$) where observations are possible:
\begin{equation}
 a(\delta,\lambda) = \int_0^{2 \pi} ~\frac{dh}{2 \, \pi}
 ~ \cos\theta_z(h, \delta, \lambda) 
 ~\Theta[\theta_{\rm max} - \theta_z(h, \delta, \lambda)]
\end{equation}
In this equation $h$ is the hour angle, $\theta_z(h, \delta, \lambda)$ is the
zenith angle of a point in the sky with declination $\delta$
as seen by a detector of latitude $\lambda$, and the factor
$\cos \theta_z$ accounts for the geometrical reduction of the
detection area when the source point has zenith angle $\theta_z$ (assuming a flat, horizontal detector).
The quantity $a(\delta, \lambda$), independently from the detector latitude,
satisfies the condition:
\begin{equation}
\int_{-1}^{1} ~d\sin \delta ~a(\delta, \lambda) = \frac{\sin^2 \theta_{\rm max}}{2} 
\end{equation}
that expresses the obvious fact that the integral of the
exposure over the entire visible part of the celestial sphere
is independent from the detector geographical position.
The top panel of Fig.~\ref{fig:detection1} shows the quantity $a(\delta,\lambda)$ as a function
of declination for some examples of the detector latitude.
It is obviously of great interest to study the gamma ray flux in the Galactic equatorial region,
in a broad interval of longitude. This is better achieved combining
the observations of more than one detector.
The bottom panel of Fig.~\ref{fig:detection1} shows the exposure factor $a(\Omega)$
for points on the Galactic equator and different detector locations.

The gamma ray signal is detected together with a
background of events generated by cosmic rays: 
\begin{equation}
B \simeq E ~\phi_{\rm cr} (E) ~\Delta \ln E ~\Delta \Omega ~A \; T \; a(\delta, \lambda)
 ~\varepsilon_{\rm cr}(E) ~.
\label{eq:back-rate}
\end{equation}
In this equation $\phi_{\rm cr}(E)$ is the cosmic ray flux, and
$\varepsilon_{\rm cr}(E)$ is the fraction of the cosmic ray showers
that survives after cuts designed to select photon showers
(for example muon multiplicity and/or structure of the shower front).

Note that in general an air shower detector will measure
the primary particle energy using an observable,
such as the shower size, that is correlated to $E$.
The correlation between the observable and the energy 
will be in general different for showers generated by gamma rays or protons/nuclei,
and it is important to take into account this difference. 
For example, if the shower size is used as an estimate of the energy,
the selection of showers in a fixed interval of size
selects photons and proton shower of different energy.
Showers generated by photons have on average a size larger
than hadronic showers of the same primary energy.
This effects reduces the background, and can
be included in the definition of the factor
$\varepsilon_{\rm cr}(E)$ of Eq.~(\ref{eq:back-rate}).

The requirement that an observable signal must be
larger than the background fluctuations ($S/\sqrt{B} \gtrsim n_\sigma$)
results in the minimum detectable flux:
\begin{equation}
 E ~\phi_{\gamma, {\rm min}} (E, \Omega) \approx
 \frac{n_\sigma}{\varepsilon_\gamma(E)}
~ \sqrt{
 \frac{E \; \phi_{\rm cr} (E) ~\varepsilon_{\rm cr}(E)}
 {A \, T \, a(\delta,\lambda) \, \Delta \Omega \, \Delta \ln E} }
\label{eq:sensitivity1}
\end{equation}
For $B \lesssim 1$, the condition for the minimum detectable flux becomes simply:
\begin{equation}
 E ~\phi_{\gamma, {\rm min}} \approx
 N_{\rm min} ~ \left [
 A \, T \, a(\delta,\lambda) \, \Delta \Omega \, \Delta \ln E \, \varepsilon_\gamma(E)
 \right ]^{-1} ~.
\label{eq:sensitivity2}
\end{equation}
where $N_{\rm min}$ is the minimum number of events for a detection.

\begin{figure}[hbt]
\begin{center}
\includegraphics[width=8.0cm]{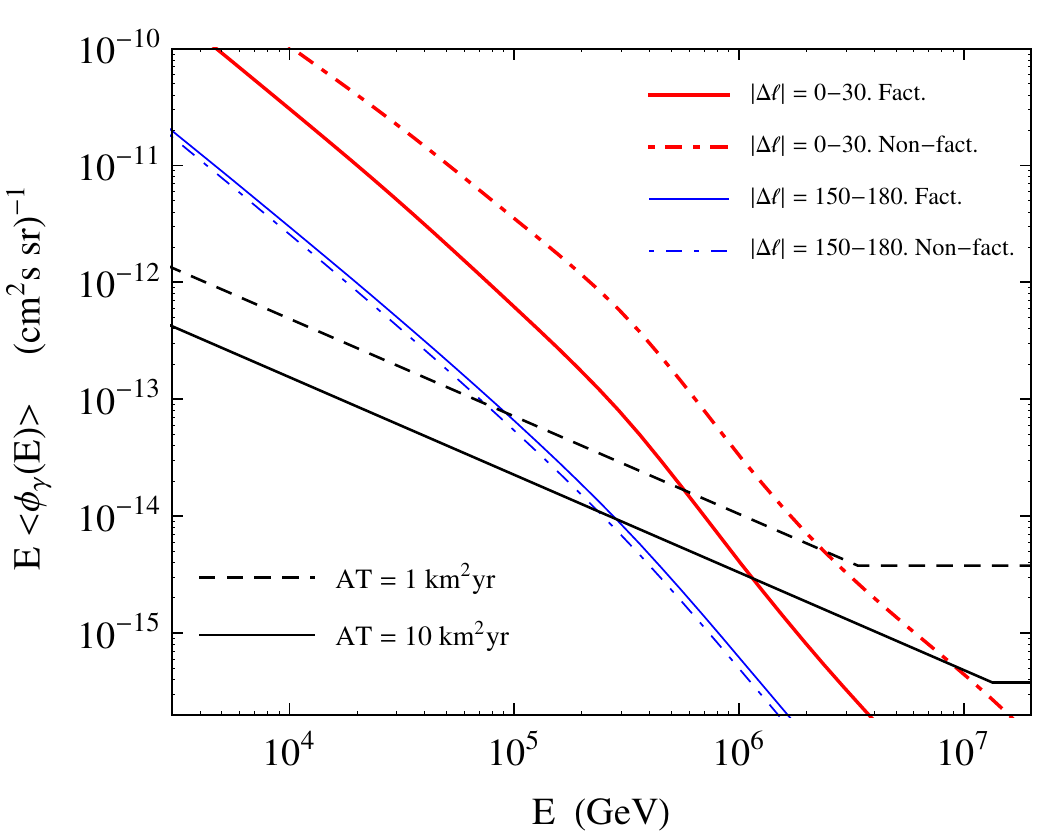}
\end{center}
\caption {\footnotesize
 Mimimum detectable diffuse gamma ray flux for an air shower detector, with
 exposure AT = 1 and 10~km$^2$~yr.
 The estimate is given for a sky region
 of size $\Delta \Omega \simeq 0.912$~sr (corresponding to
 a region of Galactic coordinates $|b| < 5^\circ$, $\Delta \ell = 30^\circ$),
 and energy bin of size $\Delta \log_{10} E \simeq 0.25$, and assuming
 $\varepsilon_{\rm cr}(E) \simeq 10^{-4}$, $\varepsilon_{\gamma}(E) \simeq 0.8$ and $n_\sigma = 3$.
 The minimum required number of events 
 was chosen as $N_{\rm min} = 10$.
 The other lines give the average gamma ray flux for two different angular 
 region of the Galaxy, according to our models (factorized and non--factorized emission).
 \label{fig:diffuse-acceptance}}
\end{figure}

In Fig.~\ref{fig:diffuse-acceptance} the sensitivity estimated with
Eqs.~(\ref{eq:sensitivity1}) and~(\ref{eq:sensitivity2}) is compared with the expected diffuse flux
estimated in the present work. One can conclude that a detector with an area
or order one km$^2$ with an hadron rejection factor of order $10^{-4}$ has the potential
to perform very interesting studies up to energies of order of several PeV's.

In this discussion we have assumed that the diffuse gamma ray signal
from the desired angular region can be estimated subtracting a background
that is measured observing other regions in the sky where the signal
is absent (or much smaller). To study (or set limits to) a diffuse flux 
that is quasi isotropic, this method cannot be applied.
In this case the identification of the gamma ray signal must rely on
a (Montecarlo based) absolute prediction of the rejection power
for hadronic showers after use of the
appropriate gamma ray selection algorithms. In this case the sensitivity
of a telescope could be limited by systematic uncertainties
in the description of hadronic cascades in the atmosphere.

\section{Outlook}
\label{sec:outlook}
The extension of the observations of the
Galactic diffuse gamma ray flux to higher energy, in the TeV and PeV
range, is a very important scientific goal that can give essential information
on Galactic cosmic rays.

In this work we have focused the discussion on the dominant contribution
to the flux, generated by the hadronic mechanism. The study of this dominant component
allows to measure the spectra of protons and nuclei in distant regions of the Galaxy, and
to determine their space dependence.
At the moment it is known that in the energy range 10--100~GeV
the CR density near the Galactic Center
is a factor two to three times larger than what is observed at the Earth, however the
question of the space dependence of the shape of the spectra remains uncertain.

The ratio between the fluxes predicted in  models where the
CR spectra have identical shape in the entire
Milky Way and in models where the spectra in the central region of the Galaxy are harder,
can become as large as one order of magnitude for gamma rays 
in the PeV energy range and directions close to the Galactic Center.
These effects can be studied by air shower arrays with sufficiently good detection capabilities
(area $A \gtrsim 1$~km$^2$ and rejection for hadronic shower $\varepsilon_{\rm cr} \lesssim 10^{-4}$).
The energy range $E \gtrsim 100$~TeV is particularly important because it allows to 
obtain information about CR in the ``knee'' region also for distant parts of the Galaxy.

Measuring the subdominant leptonic contribution to the diffuse Galactic emission
is a difficult but very important task, that can give fundamental
information on the space dependence of the $(e^+ + e^-)$ spectra.
The leptonic component could become observable in the flux at large Galactic latitudes,

The effects of absorption of high energy gamma rays propagating
over Galactic distances are important, and are largest for $E_\gamma$ of order 1-3~PeV.
The mean free path has its minumum value (of order 7~Kpc) for $E \simeq 2.2$~PeV.
Photons of this energy emitted near the Galactic Center
have a survival probability of order 0.29.
This implies that a large part of the Galactic volume is effectively unobservable
with PeV gamma rays, however a large fraction of the diffuse flux has its origin
at shorter distances, and therefore observations in the PeV range
can give very important information on the CR space distributions.
The Galactic Center itself can in fact be studied even if the flux is suppressed.

The measurement of the diffuse gamma ray flux in a
large angular region is very important for an understanding
of the space dependence of the CR spectra, of particular importance is to observe the entire
Galactic disk. Since ground based detectors can view only a fraction of the sky,
it is desirable to have multiple telescopes at different latitude.

The problem of disentangling the diffuse flux from the flux of unresolved discrete sources
is not completely solved even at low energy,
and it is likely that this question
will be more important in an energy range (0.1--10~PeV) where 
little is now known about the sources. It should however be noted 
that also a measurement of the sum of the two (diffuse and source) components
can be very useful to develop an understanding of the Galactic cosmic rays.

The same information of the CR spectra that can be inferred from the
observation of gamma rays is also contained in the spectra of
Galactic neutrinos. The simultaneous measurement of the gamma ray and neutrino fluxes
is clearly very desirable and very important to study the existence of a variety
of different effects, including the possibility of non standard mechanism of production,
or non--standard propagation for $\nu$'s and/or $\gamma$'s.

\vspace{0.25 cm}

\noindent{\bf Acknowledgments.} During the preparation of this work we benefitted from
discussion with several colleagues. We are especially grateful to
Ralph Engel,
Tom Gaisser,
Elisa Resconi,
Todor Stanev,
Francesco Villante,
Francesco Vissani
and Cao Zhen.

\end{document}